\RequirePackage{etex}

\documentclass{aa}

\usepackage[utf8]{inputenc}
\usepackage[T1]{fontenc}
\usepackage[varg]{txfonts}  % A&A required TX Times fonts
\usepackage{amsmath,amssymb}
\usepackage{graphicx}
\usepackage{natbib}
\bibpunct{(}{)}{;}{a}{}{,}  % A&A citation style
\usepackage[pdfusetitle=false,hidelinks,hypertexnames=false,unicode]{hyperref}
\hypersetup{
  pdftitle={At the stellar noise frontier: A transit survey of 121 TESS M3-M6 dwarfs},
  pdfauthor={Yohann Tschudi},
  pdfsubject={Exoplanet detection},
  pdfkeywords={exoplanets, M-dwarfs, transits, sub-Earths, planet candidates}
}

\usepackage{orcidlink}
\usepackage{xcolor}
\usepackage{booktabs}
\usepackage{multirow}
\usepackage{placeins}
\usepackage{float}
\usepackage{tikz}
\usetikzlibrary{shapes.geometric, arrows.meta, positioning, fit, backgrounds}

\title{At the stellar noise frontier: A transit survey of 121 TESS M3-M6 dwarfs}
\titlerunning{Transit survey at the noise frontier: 121 M3-M6 dwarfs}

\author{Yohann Tschudi\inst{1}\,\orcidlink{0009-0002-7797-1502}}
\authorrunning{Y. Tschudi}

\institute{Independent Researcher, 69480 Anse, France\\
\email{yohann.tschudi@gmail.com}}

\date{Submitted to Astronomy \& Astrophysics (manuscript ref. aa59754-26)}

\abstract 
{
As  the most favorable hosts for small transiting planets, M-dwarf stars can be very useful in detecting these objects. A newly accessible discovery space has recently opened up thanks to mid-to-late M-dwarfs characterized by sufficient TESS multisector coverage thanks to new Cycle~6+ observations. 
}
{This paper presents a systematic transit survey of 121 ``newly enabled'' M3-M6 dwarfs ($T_{\mathrm{eff}}=2700$--$3400$\,K) over a period of $P=0.5$-$100$\,d. These surveyed stars only recently crossed the multisector TESS detection threshold via Cycle~6+ coverage.
}
{The sample was selected from 498\,312 TIC M-dwarfs via a nine-step funnel. The pipeline combines a transit least squares approach (TLS) with a signal validation cascade, TRICERATOPS vetting, Gaia data release 3 (DR3) verification, and three empirical signal reliability tests.
}
{The pipeline validation achieved 100\% recovery (16/16 planets) on ten known systems with zero false positives. The survey identified 20 transit-like signals across 16 systems, none of which had been characterized by any prior TESS object of interest (TOI) designations. The reliability framework classified two as tier~1 (high robustness), seven as tier~2 (moderate), and ten as tier~3 (noise-susceptible), while one monotransit was excluded. For nine out of 16 hosts, the candidate SDE sits at or below the measurable noise floor. Of the remaining seven, the floor lies below $\mathrm{SDE}_{\rm inv}=7$ and the candidates end up exceeding it. The global false alarm rate is $17.4\%$ ($21/121$; Wilson 95\% CI: $[11.6\%, 25.1\%]$).
}
{The survey quantifies TLS sensitivity limits on active M3-M6 dwarfs with sparse TESS coverage. The two tier~1 candidates are priorities for RV confirmation. The ten tier~3 candidates require additional TESS sectors to establish a signal persistence, while nine systems call for high-resolution imaging to confirm their classification.
}


\begin{document}
\columnwiselinenumberstrue    % Per-column numbering: left col → left margin, right col → right margin
\switchlinenumbers            % Outer margin placement (already loaded by aa.cls)

\maketitle
\nolinenumbers

\section{Introduction}

M-dwarf stars, comprising approximately 70\% of all stars in the solar neighborhood \citep{Henry2006}, are the most favorable hosts for small transiting planets and they are useful in detecting these objects. Their small radii ($0.1$--$0.6~R_\odot$) produce deeper transit signals for a given planet size, while their proximity to the Sun enables detailed follow-up studies with current and upcoming facilities \citep{Kempton2018}. The Kepler mission established that small planets ($R_p < 1.5~R_\oplus$) are abundant around M-dwarfs, with an occurrence rate of $\sim$2.5 planets per star for $P < 200$~d \citep{Dressing2015}. However, the census of sub-Earths ($R_p < 1.0~R_\oplus$) from TESS remains sparse, as most TESS surveys of M-dwarfs are restricted to short periods ($P < 7$~d; \citealt{Ment2023}) or bright targets ($T_{\rm mag} < 10.5$; \citealt{Guerrero2021}).

The Transiting Exoplanet Survey Satellite \citep[TESS;][]{Ricker2015} has systematically observed the entire sky, with the ecliptic poles receiving continuous coverage over many years. This accumulated baseline fundamentally transforms the detection space for transiting planets: while single 27-day sectors limit searches to periods $P \lesssim 13$ days, baselines exceeding 1\,000 days enable detection of planets with periods up to $P \sim 70$--$100$ days ($P_{\rm max}=0.7\,B$, where $B$ is the total observational baseline in days), probing a parameter space inaccessible to single-sector surveys.

With TESS Cycle~6+ observations (Sectors 80+, 2024--2025), a new population of M3--M6 dwarfs has crossed the multisector detection threshold: stars that had $\leq$2 archival sectors and only recently acquired sufficient coverage for transit searches at extended periods. Previous large-scale surveys focused on bright stars amenable to radial velocity follow-up \citep{Guerrero2021} or employed reliability-optimized thresholds that prioritize high-confidence candidates at the expense of completeness in the small-planet regime \citep{Thompson2018}. This newly available coverage can be exploited to conduct a deliberate test of TLS detection limits on active mid-to-late M-dwarfs, while probing the stellar noise frontier cited in the survey's name.

This paper presents a systematic transit survey of 121 M-dwarf stars spanning spectral types M3--M6 ($T_{\mathrm{eff}} \approx 2\,700$--$3\,400$~K), specifically selected as ``newly enabled'' targets: stars that only became searchable through the accumulation of recent TESS sectors, with confirmed single-star status from Gaia data release 3 (DR3) astrometry. The survey covers the full accessible period range ($P=0.5$--$100$~d). The sample includes two M6 dwarfs (deepest transits, $\sim$4000~ppm for Earth-sized planets), 22 M4--M5 dwarfs (moderate depths, extended periods accessible with typical baselines), and 97 M3 dwarfs (shallowest transits, longest accessible periods).

A pipeline was developed and validated based on a transit least squares approach \citep[TLS;][]{Hippke2019} with Scout and Sniper optimization to improve the computational efficiency on long baselines, enabling a systematic exploration of the period-radius parameter space with false-positive rejection through multistage vetting with TRICERATOPS \citep[Tool for Rating Interesting Candidate Exoplanets and Reliability Analysis of Transits Originating from Proximate Stars;][]{Giacalone2021}. While TLS provides a detection threshold (SDE~$\geq 7$), the false alarm rate on active M-dwarfs with sparse multisector coverage remains poorly characterized in an empirical sense. This work addresses this gap by combining detection with a systematic signal reliability assessment based on three independent empirical tests (circular shift FAR, light-curve inversion, and Fourier phase scrambling). The approach presented here differs from existing TESS pipelines in several key aspects (Table~\ref{tab:comparison}):

\begin{table*}[!htbp]
\centering
\small
\caption{Comparison of transit detection pipelines.}
\label{tab:comparison}
\begin{tabular}{lcccc}
\toprule
\textbf{Aspect} & \textbf{SPOC/QLP} & \textbf{SHERLOCK}$^a$ & \textbf{Nigraha}$^b$ & \textbf{This work} \\
\midrule
Algorithm & BLS & TLS/BLS & TLS + ML & TLS \\
Detrending & PDC & Savitzky-Golay & Standard & GP + Wotan \\
Multi-sector & Per-sector & Yes & Per-sector & Stitched (3--10) \\
Validation & TOI alerts & Statistical & Neural net & TRICERATOPS \\
Specificity & All stars & General & Shallow transits & M3--M6 noise frontier \\
\bottomrule
\end{tabular}
\tablefoot{BLS=box least squares; TLS=transit least squares; ML=machine learning; PDC=pre-search data conditioning; GP=Gaussian process; SPOC=science processing operations center; QLP=quick look pipeline. TRICERATOPS is defined in Sect.~\ref{sec:vetting}. $^a$\citet{DevorasPajares2024}. $^b$\citet{Rao2021}.}
\end{table*}

\noindent While each individual component is characterized by specific precedents (TLS, TRICERATOPS, GP detrending, etc.), the combination targets a specific niche: (1) extended period search ($P_{\max}=0.7B$, reaching 50--100~d) via multisector stitching, extending searchable periods from $P \lesssim 7$~d (single sector) to $P \sim 50$--$100$~d; (2) 18-check signal validation cascade targeting M-dwarf-specific false positives (BY~Dra variability, harmonic nests); and (3) Gaia~DR3 physical verification systematically integrated into the pipeline for direct background eclipsing binary exclusion.

The remainder of this paper is organized as follows: target selection (Sect.~\ref{sec:sample}), detection and validation methodology, including the signal reliability framework (Sect.~\ref{sec:methodology}), injection-recovery completeness tests (Sect.~\ref{sec:injection}), pipeline robustness on known systems (Sect.~\ref{sec:validation}), limitations (Sect.~\ref{sec:limitations}), results including signal reliability assessment (Sect.~\ref{sec:results}), discussion (Sect.~\ref{sec:discussion}), and conclusions (Sect.~\ref{sec:conclusions}).

\section{Target selection strategy}
\label{sec:sample}

\subsection{The ``newly enabled M3-M6'' sample: Nine-step selection funnel}

The central challenge in designing this survey was identifying M-dwarfs where making transit detections at extended periods has only recently become possible, as stars had previously been overlooked by existing surveys due to insufficient temporal coverage. For M3-M6 dwarfs, multisector baselines of several hundred to $>$2\,500~d extend the searchable period range to $P \sim 50$--$100$~d, well beyond the reach of single-sector surveys. To isolate the target population, a nine-step selection funnel was applied to the 498\,312 TIC M-dwarfs, progressively narrowing to stars that crossed the multisector detectability threshold with TESS Cycle~6+ observations. 

\begin{table*}[!htbp]
\centering
\caption{Newly enabled M3-M6 selection funnel.}
\label{tab:funnel}
\begin{tabular}{@{}cp{0.38\textwidth}rp{0.25\textwidth}@{}}
\toprule
Step & Criterion & Targets & Justification \\
\midrule
1 & TIC M-dwarfs ($T_{\mathrm{eff}} \leq 4000$~K) & 498\,312 & Complete M-dwarf base \\
2 & $T_{\mathrm{eff}}=2700$--$3400$~K (M3--M6) & 114\,107 & Deep transits ($>$800~ppm for $1~R_\oplus$) \\
3 & $R_\star < 0.35~R_\odot$ & 59\,414 & Depth $>$1000~ppm for $1~R_\oplus$ \\
4 & $T_{\rm mag}=8.0$--$12.5$ & 1\,852 & Avoids saturation and faint noise \\
5 & Contamination ratio $< 0.05$ & 1\,475 & Dilution $<$5\% \\
6 & $\geq$3 total sectors & 1\,061 & Period disambiguation + S/N\\
7 & Newly enabled$^a$: $\leq$2 archival + $\geq$1 Cycle~6 + $B > 100$~d & 199 & \textbf{Not searchable before Cycle~6} \\
8 & Gaia RUWE $< 1.4$ & 174 & Excludes unresolved binaries \\
9 & No TOI or known planet & \textbf{121} & Pure discovery targets \\
\bottomrule
\end{tabular}
\tablefoot{Counts based on TIC v8.2 cross-matched with TESS observation logs through Sector~97 and Gaia DR3. All counts are exact, derived from TIC v8.2 (steps 1--5) and \texttt{lightkurve} sector-level cross-match through Sector~97 (steps 6--7). S/N=signal-to-noise ratio. RUWE=renormalised unit weight error. TOI=TESS object of interest. $^a$The three subcriteria are logically linked: fewer than two archival sectors combined with more than three total guarantees of more than one recent sector and $B > 100$~d.}
\end{table*}

\subsubsection{Sample composition by spectral type}

The sample is restricted to spectral types M3--M6 ($T_{\mathrm{eff}}=2700$--$3400$~K), excluding earlier M-dwarfs (M0--M2) for two reasons: (1)~for M0--M2 dwarfs ($T_{\mathrm{eff}} > 3400$~K, $R_\star \approx 0.4$--$0.6~R_\odot$), Earth-sized transits produce $\delta \approx 300$--$600$~ppm, approaching the detection threshold (Signal Detection Efficiency, SDE~$\geq 7$; \citealt{Hippke2019}) and increasing false positive rates; meanwhile (2)~the ``newly enabled'' selection, by construction, has limited archival data, so multisector coverage is insufficient for extended-period searches on M0--M2 dwarfs in this sample.

Table~\ref{tab:sample_types} summarizes the sample composition. The M3-M6 range provides complementary detection sensitivities:
\begin{itemize}
\item M6 dwarfs ($R_\star \approx 0.15~R_\odot$, $T_{\mathrm{eff}}=2700$--$2900$~K): deepest transits ($\delta \approx 3\,700$~ppm for $1~R_\oplus$), enabling sub-Earth detection even with sparse coverage, but rare in magnitude-limited samples (N~$= 2$);
\item M4--M5 dwarfs ($R_\star \approx 0.20$--$0.25~R_\odot$, $T_{\mathrm{eff}}=2900$--$3200$~K): the so-called sweet spot combining moderate transit depths ($\delta \approx 1\,300$--$2\,100$~ppm for $1~R_\oplus$) with extended-period sensitivity well-matched to the survey baselines (N~$= 22$);
\item M3 dwarfs ($R_\star \approx 0.30$--$0.35~R_\odot$, $T_{\mathrm{eff}}=3200$--$3400$~K): shallower transits ($\delta \approx 700$--$950$~ppm for $1~R_\oplus$) requiring a higher signal-to-noise ratio (S/N) and longer baselines, while representing the largest subsample (N~$= 97$).
\end{itemize}

\begin{table}[!htbp]
\centering
\caption{Sample distribution by spectral type.}
\label{tab:sample_types}
\begin{tabular}{lcc}
\toprule
Type & $T_{\mathrm{eff}}$ (K) & N \\
\midrule
M6 & 2700--2900 & 2 \\
M4--M5 & 2900--3200 & 22 \\
M3 & 3200--3400 & 97 \\
\midrule
\textbf{Total} & 2700--3400 & \textbf{121} \\
\bottomrule
\end{tabular}
\end{table}

\subsection{Selection criteria}

Each step in Table~\ref{tab:funnel} is motivated by specific scientific or technical requirements. The rationale for each criterion is detailed below.

\begin{itemize}
\item Steps 1-2: spectral type restriction ($T_{\mathrm{eff}}=2700$--$3400$~K). The lower bound ($T_{\mathrm{eff}} \geq 2700$~K) excludes late M-dwarfs with poorly constrained stellar parameters. The upper bound ($T_{\mathrm{eff}} \leq 3400$~K) excludes M0--M2 dwarfs where transit depths for Earth-sized planets fall below the detection threshold and multisector coverage is insufficient for extended-period searches (see Sect.~2.1.1).
\item Step 3: stellar radius ($R_\star < 0.35~R_\odot$). This ensures transit depths $\delta > 800$~ppm for Earth-sized planets, well above the detection limit. Larger radii produce shallower transits that approach the noise floor.
\item Step 4: magnitude range ($T_{\rm mag}=8.0$--$12.5$). The lower limit excludes ultra-bright stars with severe instrumental red noise ($\beta > 5$), which prevents reliable transit detection even for known planets (see Sect.~\ref{sec:limitations}). The upper limit ensures sufficient photometric precision ($\sigma < 1000$~ppm/hr) for Earth-sized transit detection.
\item Step 5: contamination ratio ($< 0.05$): nearby sources contributing $>$5\% flux dilute transit depths and introduce false positive scenarios. This threshold ensures that even 1000~ppm transits remain detectable after dilution correction.
\item Step 6: minimum sectors ($\geq 3$). Multiple sectors provide (1)~period disambiguation through phase coherence, (2)~S/N boost from transit stacking, and (3)~harmonic rejection via transit count verification.
\item Step 7: ``newly enabled'' criterion ($\leq 2$ archival sectors, $\geq 1$ Cycle~6 sector, baseline $B > 100$~d). This is the central innovation of this survey. The strategy targets stars that were not previously searchable at extended periods (insufficient archival coverage) but became searchable with recent TESS observations (Cycle 6+, Sectors 80+). Planets at periods $P=10$--$100$~d require multisector coverage to accumulate $\geq 3$ transits for reliable period determination. With three sectors spread across TESS cycles, $B$ can reach 700--2\,600~days despite only $\sim$80~days of actual photometry (Table~\ref{tab:sample_summary}). This criterion isolates genuinely unexplored discovery space, avoiding targets already examined by archival mining surveys.
\item Step 8: Gaia renormalised unit weight error (RUWE) ($< 1.4$). The standard threshold from \citet{Lindegren2021} is adopted to exclude unresolved binaries. Elevated RUWE indicates excess astrometric scatter from unresolved companions, which dilute transit depths and complicate interpretation.
\item Step 9: no known planets or TESS objects of interest (TOIs). Targets with existing TOI alerts or confirmed planets are excluded. Together with Step~7, this restriction defines a discovery-only sample: every detected signal corresponds to a newly identified candidate, and the false-alarm rate and detection-sensitivity statistics reported in Sects.~\ref{sec:tier} and \ref{sec:far} characterize the intrinsic behavior of TLS on active mid-to-late M-dwarfs, independently of the selection imposed by other transit pipelines.
\end{itemize}

The final sample comprises 121 M-dwarf stars representing a well-defined, previously unexplored parameter space for transit detection around mid-to-late M-dwarfs. Based on the defined sample, the following section describes the methodology employed to search these targets for transiting planets.

\section{Methodology}
\label{sec:methodology}

Detecting small planets around faint M-dwarfs requires careful attention to both signal extraction and false-positive rejection. The pipeline processes TESS photometry through six sequential stages, each addressing specific challenges: (1)~data acquisition via quality cascade (SPOC~$\rightarrow$~TESS-SPOC~$\rightarrow$~QLP) to maximize photometric precision; (2)~Gaussian Process (GP)-based preprocessing and sector stitching to remove stellar variability while preserving transit shapes; (3)~TLS transit detection with SDE~$\geq 7$ threshold to identify candidate signals; (3b)~harmonic nest cleanup via cross-masking and system parsimony for multicandidate systems; (4)~TRICERATOPS statistical vetting to quantify false-positive probabilities; and (5)~Gaia~DR3 physical verification to directly test background eclipsing binary scenarios.

\subsection{Data acquisition and preprocessing}
\label{sec:acquisition}

TESS photometry was retrieved using \texttt{lightkurve} \citep{Lightkurve2018} with a quality cascade prioritizing SPOC (2 min) over TESS-SPOC over QLP (30 min). Each sector was normalized by median flux with asymmetric sigma-clipping ($4\sigma$ above, $20\sigma$ below to preserve deep transits).

Stellar parameters ($R_\star$, $M_\star$, $T_{\mathrm{eff}}$) were adopted directly from TIC~v8.2 \citep{Stassun2019} without applying specialized M-dwarf relations \citep[e.g.,][]{Mann2015}. Distances are cross-validated with Gaia~DR3 parallaxes \citep{GaiaDR3}. For mid-M dwarfs, TIC radii carry systematic uncertainties of 10--20\% \citep{Stassun2019}, propagating directly to planetary radii as $\sigma_{R_p}/R_p \approx \sigma_{R_\star}/R_\star$. Consequently, a candidate reported as $R_p=0.8~R_\oplus$ might range from $0.65$--$0.95~R_\oplus$. Spectroscopic refinement of $R_\star$ is required before definitive sub-Earth classification. Spectral type assignments (M3-M6) are based on TIC $T_{\mathrm{eff}}$; uncertainties of $\pm$100--200~K can shift the classification by one subtype.

\label{sec:preprocessing}
Detrending uses Gaussian process regression via \texttt{celerite2} \citep{ForemanMackey2017} with a two-component kernel: a stochastic harmonic oscillator (SHOTerm, quasi-periodic) for stellar variability and a Mat\'ern-3/2 term for long-term trends. The primary timescale, $\rho$, is adaptive ($0.3$--$5.0$~d) with a hard floor at $0.3$~d to preserve transit signals. For GP convergence failures, the pipeline falls back to biweight detrending via \texttt{wotan} \citep{Hippke2019wotan}. Stellar flares (common in M-dwarfs) are handled through the asymmetric sigma-clipping ($4\sigma$ above median): isolated positive excursions are clipped while preserving the symmetric transit dips. The GP's long timescale ($>0.3$~d) further prevents flare contamination of the baseline, as flares typically last minutes to hours. Detrended sectors are stitched chronologically, yielding multisector light curves spanning 700--2\,600~days with $\sim$50\,000--200\,000 measurements per target.

\subsection{Transit detection}
\label{sec:search}

Transit detections use TLS \citep{Hippke2019} with quadratic limb-darkening coefficients computed for the TESS bandpass from \citet{Claret2017}. Key parameters: period range $0.5 < P < \min(0.7B, 100)$~days, with a detection threshold of SDE$_{\rm eff} \geq 7.0$ and an iterative search for up to six planets per system. The effective SDE is computed as SDE$_{\rm eff}$=SDE$_{\rm TLS}$~$/\sqrt{\beta_{\rm rn}}$, where SDE$_{\rm TLS}$ is the raw output of the \texttt{transitleastsquares} library and $\beta_{\rm rn}=(\sigma_{\rm binned}/\sigma_{\rm white})^2 \geq 1$ is the red-noise variance ratio of \citet{Pont2006} that accounts for time-correlated noise. This penalization is applied as a custom post-processing step to the standard TLS output. All subsequent SDE values in this paper refer to SDE$_{\rm eff}$.

For long baselines ($B > 500$~d), a two-phase Scout+Sniper strategy reduces computation by a factor scaling with $N_{\rm raw}/N_{\rm bin}$ (typically 2--5$\times$ for the survey baselines): Scout phase bins the light curve (reducing $N$ by ${\sim}5\times$) and searches over the full period grid with a relaxed threshold (SDE~$\geq 4.5$); Sniper phase refines only the Scout detections on unbinned data within $\pm 10\%$ of Scout periods. The Scout threshold of 4.5 was set empirically so that all 16 known planets of the validation sample (Sect.~\ref{sec:validation}) could be recovered between the Scout and Sniper phases. The $\sim$2.5-SDE margin below the SDE~$\geq 7$ detection requirement accommodates the S/N loss incurred by binning; marginal losses on targets outside the validation sample remain possible and are discussed in Sect.~\ref{sec:limitations}. The speedup scales with $N_{\rm raw}/N_{\rm bin}$, since TLS computation is $\mathcal{O}(N \times M)$, where $M$ is the period grid size.

\subsubsection{Harmonic correction}
\label{sec:subharmonic}

Harmonic aliasing occurs when TLS detects $2P$ or $3P$ instead of the true period, $P$. Three evidence modes address this event, each targeting a distinct aliasing signature:

\paragraph{Mode~A (equal-depth secondary):} inspired by the Kepler Model-Shift Uniqueness Test \citep{Twicken2018} and the Robovetter \citep{Coughlin2016,Thompson2018}, a per-epoch matched filter measures event depth at phase~0.5 using only epochs with five or more TESS data points and the mean in-transit flux within $\pm T_{\rm dur}/2$. A depth ratio exceeding~0.4 identifies a $2P$ alias, since no planetary secondary exceeds ${\sim}2\%$ of the primary depth \citep{Santerne2013}. Mode~B (odd-even depth mismatch): if $|\delta_{\rm odd} - \delta_{\rm even}| > 3\sigma$ \citep{Batalha2013}, alternating transit depths indicate a $P/2$ alias of an eclipsing binary (the detected period is half the true orbital period). Mode~C (equal-depth thirds): equal-depth events at phases~$1/3$ and~$2/3$ (ratio~$>0.4$) identified $3P$ aliases; no single planet produces three equidistant transit-like events at phases 0, 1/3, and 2/3 with comparable depths. For depth ratios~$>0.8$ (Modes~A/C), the S/N requirement is waived following the Kepler Robovetter \citep{Thompson2018}; for ratios 0.4--0.8, S/N~$>3$ is required. The pipeline resolves the period by re-running TLS at candidate subharmonics ($P/2$, $P/3$, $P/4$) with an evidence-dependent adaptive threshold:
\begin{itemize}
    \item Mode~A (equal-depth secondary, $2P$ alias): SDE~$>0.3\times$SDE$_{\rm orig}$;
    \item Mode~B (odd-even $>3\sigma$, $P/2$ alias): SDE~$>0.8\times$SDE$_{\rm orig}$;
    \item Mode~C (equal-depth thirds, $3P$ alias): SDE~$>0.3\times$SDE$_{\rm orig}$;
    \item No physical evidence: no resolution attempted (SDE ratio alone is not diagnostic).
\end{itemize}

\paragraph{Transit count correlation:} if $P_2/P_1 \approx N$ and $N_{\rm tr}(P_1)/N_{\rm tr}(P_2) \approx N$ within 10\%, the longer period is flagged as harmonic. To preserve true 2:1 mean motion resonance (MMR) systems \citep{Fabrycky2014}, the algorithm does not resolve to a period already accepted for the same target.

\subsection{Signal validation cascade}
\label{sec:signal_validation}

Each TLS detection undergoes a cascade of 18 diagnostic checks (13 REJECT, 1 METRIC, 4 LOG) within the detection module before reaching statistical vetting. Table~\ref{tab:checks} summarizes these checks; the key diagnostics are described below.

\begin{table}[!htbp]
\centering
\caption{Signal validation cascade: 18 diagnostic checks.}
\label{tab:checks}
\footnotesize
\setlength{\tabcolsep}{3pt}
\begin{tabular}{@{}clcl@{}}
\toprule
\# & Check & Type & Threshold \\
\midrule
1 & Signal polarity & R & S/N~$\leq 0$\\
2 & Period aliases & R & $P \approx$ 0.5, 1.0, 13.7~d \\
3 & Sub-harmonics & R & $P/n$ of accepted cand. \\
4 & Transit count corr. & R & $N_{\rm tr}$ ratio $\approx N$ \\
5 & Dynamical stability & R & $\Delta < 3.5\,R_{\rm Hill}$ \\
6 & Centroid offset & R & $> 5\sigma$ \\
7 & Secondary eclipse & R & Depth ratio $> 0.10$ \\
8 & Odd-even depth & R & $> 3\sigma$ \\
9 & Duration plausibility & R & $T_{\rm obs}/T_{\rm exp} \notin [0.3, 3.0]$ \\
10 & NTL spike test & R & Median/mean $< 0.20$ \\
11 & SWEET variability & R & $A_{\rm ratio} < f(\delta)$ \\
12 & OOT scatter gate & R & MAD $> 2\times$ expected \\
13 & Detrend.\ response & R & Pre/post-GP $> 2.5$ \\
14 & Depth consistency & M & Four-layer diagnostic\\
15--18 & Transit parameters & L & Logged for catalog \\
\bottomrule
\end{tabular}
\tablefoot{R=REJECT (13 checks), M=METRIC (1; recorded but not rejection), L=LOG (4; audit trail). Checks are applied sequentially; a REJECT terminates the cascade for the current candidate.}
\end{table}

\begin{itemize}
\item Signal polarity (check~1): candidates with S/N~$\leq 0$ are rejected as inverted signals, which would be physically impossible for a transit, as it must produce a flux decrease. On active M-dwarfs (MAD~$> 2500$~ppm), correlated stellar variability can inflate SDE while producing negative S/N (upward bump); this check fires before all other diagnostics.
\item Instrumental rejection (checks~2 and~10): candidates at known systematic periods are rejected: TESS orbital period ($13.7 \pm 2\%$~d), Earth rotation aliases ($0.5$ and $1.0$~d within $\pm 2\%$; \citealt{Fausnaugh2018}). A spike test inspired by the Kepler Robovetter not-transit-like (NTL) metric \citep{Coughlin2016,Thompson2018} rejects instrumental artifacts (momentum dumps, cosmic rays) when the ratio of median to mean in-transit depth falls below~$0.20$ ($\geq 4$ in-transit points required): real transits depress all in-transit points uniformly (ratio~$\approx 1$), while isolated outliers affect only the mean (ratio~$\approx 0$).
\item Eclipsing binary rejection (Checks~5--8): odd-even depth comparison ($>3\sigma$) rejects EBs \citep{Batalha2013}. Centroid analysis ($>5\sigma$ offset) rejects blends \citep{Bryson2013}. Secondary eclipse detection (depth ratio~$>0.10$) rejects EBs, with the S/N requirement waived for ratio~$>0.15$ \citep{Santerne2013}; for multiplanet candidates with near-2:1 period ratios, the secondary rejection is bypassed when $P/2$ matches an accepted candidate (within 3\%), preventing false rejection from transit masking artifacts. Dynamical stability is checked for multiplanet systems ($\Delta > 3.5$ mutual Hill radii; \citealt{Gladman1993,Chambers1996}).
\item Stellar rotation estimation (input to check~14): rotation periods are estimated via the Generalised Lomb-Scargle periodogram \citep{Zechmeister2009} on pre-GP flux (FAP~$<5\%$), with Earth rotation and TESS orbital aliases vetoed. The relaxed FAP (from 0.1\%) maximizes $P_{\rm rot}$ sensitivity on active M-dwarfs; false assignments are mitigated by requiring phase coherence~$< 0.75$ before rejection.
\item Stellar variability rejection (Checks~11--13): short-period candidates ($P < 2\,P_{\rm rot}$, or $P < 15$~d when $P_{\rm rot}$ is unavailable) are tested through three complementary diagnostics on pre-GP flux: (1)~a sinusoidal variability test \citep[SWEET, after the Kepler Robovetter;][]{Thompson2018,Kunimoto2025} fits sinusoids at $P$, $P/2$, $2P$ per TESS segment, rejecting when the amplitude ratio $A_{\rm ratio}=\delta / (2\,A_{\rm sine,\,median})$ falls below a depth-scaled threshold $\max(1.5,\, 2.5\,(1 - e^{-\delta/1000}))$; (2)~a Folded OOT Scatter Gate rejects when binned out-of-transit MAD exceeds both $2\times$ expected noise and $30\%$ of depth; and (3)~a Detrending Response Test rejects when depth ratio pre-GP/post-GP~$> 2.5$, indicating variability absorbed by GP.
\item Depth consistency (check~14): phase coherence is defined here as the ratio of the median per-epoch transit depth to the TLS-fit depth. A geometric transit produces consistent per-epoch depths and a ratio that approaches unity; stellar variability across epochs (spot-pattern evolution, correlated noise) depresses the median below the TLS-fit value and yields lower ratios, motivating the thresholds used by the four layers below. A four-layer diagnostic addresses variability at all periods: Layer~1 rejects phase coherence~$< 0.5$; Layer~2 rejects inter-sector depth coefficient of variation (CV)~$> 0.5$; Layer~3 rejects $P \approx P_{\rm rot}$ harmonics ($P_{\rm rot}/3$, $P_{\rm rot}/2$, $P_{\rm rot}$, $2P_{\rm rot}$, $3P_{\rm rot}$) with phase coherence~$< 0.75$; Layer~4 rejects sector CV~$> 0.70$ combined with phase coherence~$< 0.80$. Layers~1--2 are S/N-gated (bypassed when per-epoch S/N~$< 2.0$).
\item Quality flag system: following the Kepler Robovetter \citep{Thompson2018} and its TESS adaptation LEO-Vetter \citep{Kunimoto2025}, candidates receive a quality flag: \texttt{HIGH\_CONFIDENCE} (default) or \texttt{MARGINAL\_ACTIVE\_STAR} (median absolute deviation MAD~$> 5\,000$~ppm, planet number~$> 1$, SDE~$< 20$). Marginal candidates bypass statistical vetting but are retained for human review.
\item Harmonic nest cleanup: on active M-dwarfs, TLS can produce multiple false candidates per star from BY~Dra variability harmonics. A post-detection cleanup analyzes all candidates per target simultaneously via two tests: (1)~cross-masking \citep{Thompson2018,Coughlin2016}---for each candidate pair (A, B) with SDE$_A >$ SDE$_B$, the transits of A are masked ($\pm 1.5 \times T_{\rm dur}/2$) and the residual depth of B is measured; if residual depth~$< 50\%$ of the original, B is flagged as dependent on A; (2)~system parsimony, whereby the maximum number of plausible detections is capped by stellar activity (MAD~$< 1500$~ppm: 5; $1500$--$3000$: 3; $3000$--$5000$: 2; $> 5000$: 1), with a $+2$ bonus when best SDE~$> 20$ and S/N~$> 5$. These thresholds were calibrated empirically on the ten validation systems (Sect.~\ref{sec:validation}): L~98-59 (MAD $= 1\,674$~ppm, three planets) passes within the cap; TOI-700 (MAD $= 3\,615$~ppm, four planets) is protected by the SDE~$> 18$ safety guard. Cross-masking evidence takes priority over parsimony. Systems with SDE~$> 18$ or S/N~$> 7$ bypass the parsimony test to protect genuine multiplanet systems.
\end{itemize}

\subsection{Candidate validation}
\label{sec:validation_axes}

Candidates passing the signal validation cascade were assessed along two complementary axes: astrophysical validation (described in this section) to determine whether the signal is consistent with a planetary transit, rather than a false positive; whereas the signal reliability (Sect.~\ref{sec:tier}) quantifies whether the detection is statistically distinguishable from the host star's noise floor. At the noise frontier, signal reliability is the primary classification driver.

\subsubsection{Statistical vetting and parameter fitting}
\label{sec:vetting}

Candidates passing the signal validation cascade undergo false-positive screening via TRICERATOPS \citep{Giacalone2020,Giacalone2021}, which computes the false positive probability (FPP) and nearby false positive probability (NFPP). TRICERATOPS evaluates astrophysical false-positive scenarios (blends, BEBs, eclipsing binaries) but does not assess instrumental systematics or stellar variability, which are the primary contaminants on active M-dwarfs. This motivates the 18-check signal validation cascade (Sect.~\ref{sec:signal_validation}) applied before TRICERATOPS. Classification thresholds follow \citet{Giacalone2021}: FPP~$<1.5\%$ and NFPP~$<0.1\%$ for statistical validation; FPP~$\geq 50\%$ or NFPP~$> 80\%$ for rejection. Intermediate cases with clean observational diagnostics proceed to Gaia physical verification (Sect.~\ref{sec:physical_verification}). Table~\ref{tab:dispositions} lists all disposition categories.

Candidates passing vetting undergo Markov chain Monte Carlo (MCMC) parameter fitting using \texttt{emcee} \citep{ForemanMackey2013} with \texttt{batman} transit models \citep{Kreidberg2015}. Three parameters are sampled with uniform priors: the planet-to-star radius ratio, $R_p/R_\star$ on $(0, 0.5]$, the impact parameter $b$ on $[0, 1.5]$ (values $b > 1$ correspond to grazing or non-transiting configurations; the extended upper bound allows the posterior to explore the grazing boundary rather than piling up at $b=1$), and a mid-transit-time offset $\Delta T_0$ on $(-0.1\,P, +0.1\,P)$ relative to the TLS ephemeris. The orbital period, $P$, is held fixed at the TLS best-fit value: the dense TLS period grid provides the period estimate and MCMC then refines the transit-shape parameters at that period. The scaled orbital distance, $a/R_\star$, can becomputed from Kepler's third law with uncertainties propagated from $R_\star$ and $M_\star$. Quadratic limb-darkening coefficients $(u_1, u_2)$ are held fixed at the \citet{Claret2017} values for the TESS bandpass. The sampler uses 32 walkers with 1\,000 burn-in and 3\,000 production steps and eccentricity is fixed at zero. Table~\ref{tab:dispositions} summarizes all disposition categories assigned across the three validation phases.

\begin{table}[!htbp]
\centering
\caption{Pipeline disposition categories across validation phases.}
\label{tab:dispositions}
\footnotesize
\setlength{\tabcolsep}{3pt}
\begin{tabular}{lp{5.5cm}}
\toprule
Disposition & Criteria \\
\midrule
\multicolumn{2}{l}{\textit{Statistical vetting (Sect.~\ref{sec:vetting})}} \\
\texttt{LIKELY\_PLANET} & FPP $<50\%$: probable planet \\
\texttt{PENDING\_VERIF.}$^a$ & NFPP 10--80\%, observations clean \\
\texttt{GRAZING\_CAND.} & Impact parameter $b>0.9$ \\
\texttt{LIKELY\_EB} & Secondary ratio $>0.10$, prob\_EB $>15\%$, or duration anomaly \\
\texttt{AMBIGUOUS} & FPP $\geq 50\%$, NFPP $<10\%$ \\
\texttt{MARGINAL}$^b$ & MAD $>5000$~ppm, pl.\# $>1$, SDE $<20$ \\
\texttt{FALSE\_POS.} & NFPP $\geq 80\%$ or centroid $>5\sigma$ \\
\midrule
\multicolumn{2}{l}{\textit{Signal reliability (Sect.~\ref{sec:tier})}} \\
\texttt{Tier} & Signal reliability tier (1/2/3) \\
\midrule
\multicolumn{2}{l}{\textit{Gaia physical verification (Sect.~\ref{sec:physical_verification})}} \\
\texttt{PLANET\_CAND.} & 0 contaminants, RUWE $<1.4$ \\
\texttt{PC\_HEB\_CAUT.} & 0 contaminants, RUWE $\geq 1.4$ \\
\texttt{NEEDS\_HR\_IMG} & $\geq 1$ contaminant, RUWE $<1.4$ \\
\texttt{FALSE\_POS.} & $\geq 1$ contaminant, RUWE $\geq 1.4$ \\
\bottomrule
\end{tabular}
\tablefoot{\texttt{PENDING} and \texttt{LIKELY\_PLANET} candidates proceed to Gaia verification. $^a$\texttt{PENDING\_PHYSICAL\_VERIFICATION}: FPP~$\geq 50\%$ but observations clean; Gaia can exclude BEB at $\rho > 0.4''$ and upgrade the candidate. $^b$\texttt{MARGINAL\_ACTIVE\_STAR}; bypasses TRICERATOPS.}
\end{table}

\subsubsection{Physical verification via Gaia DR3}
\label{sec:physical_verification}

Background eclipsing binaries (BEB) are the dominant false-positive scenario for faint M-dwarf hosts. Gaia~DR3 directly tests BEB hypotheses (Fig.~\ref{fig:decision_tree}) by querying sources within 30\arcsec\ brighter than $R_{p,\rm max}$, the faintest contaminant capable of producing the observed depth, $\delta$:
\begin{equation}
R_{p,\rm max}=T_{\rm mag} - 2.5 \log_{10}(\delta)
\label{eq:gmax},
\end{equation}
where the contaminant search uses Gaia~$R_p$-band photometry (630--1000~nm) rather than the broadband $G$ magnitude, since Gaia~$R_p$ closely matches the TESS bandpass (600--1000~nm) for M-dwarf spectral energy distributions \citep{Fabricius2021}. For mid-M dwarfs, $T_{\rm mag}$ and Gaia~$R_p$ agree to within $\sim$0.1--0.2~mag, making $T_{\rm mag}$ a suitable proxy; the residual offset is anti-conservative (slightly underestimating $R_{p,\rm max}$), but smaller than the typical magnitude margin between the brightest contaminant and the threshold (Table~\ref{tab:app_hr_facilities}). Zero contaminants with RUWE~$<1.4$ \citep{Lindegren2021} yields \texttt{PLANET\_CANDIDATE}; $\geq$1 contaminant yields \texttt{NEEDS\_HR\_IMAGING}. Only \texttt{PENDING} and \texttt{LIKELY\_PLANET} candidates enter verification. For \texttt{LIKELY\_PLANET} candidates (FPP~$< 50\%$), TRICERATOPS has already integrated BEB priors via TRILEGAL; Gaia contaminants therefore do not downgrade these candidates but prompt a recommendation for high-resolution imaging. Gaia cannot resolve companions $<$0.4\arcsec and, thus, HR imaging is required for a confirmation.

\begin{figure}[!htbp]
\centering
\resizebox{\columnwidth}{!}{%
\begin{tikzpicture}[
    decision/.style={diamond, draw=black!60, line width=0.5pt, aspect=2.5,
                     inner sep=1pt, align=center, fill=white, font=\scriptsize},
    outcome/.style={rectangle, draw=black!70, line width=0.6pt, rounded corners=2pt,
                    minimum width=1.8cm, minimum height=0.8cm, align=center, font=\scriptsize\bfseries},
    validated/.style={outcome, fill=green!15, draw=green!50!black},
    caution/.style={outcome, fill=orange!15, draw=orange!60!black},
    pending/.style={outcome, fill=yellow!20, draw=yellow!60!black},
    rejected/.style={outcome, fill=red!10, draw=red!50!black},
    arrow/.style={-{Stealth[length=2mm, width=1.5mm]}, line width=0.5pt, draw=black!50},
    yeslabel/.style={font=\tiny, text=black!70, fill=white, inner sep=1pt},
    nolabel/.style={font=\tiny, text=black!70, fill=white, inner sep=1pt}
]

\node[font=\small, align=center] (input) at (0,0) {Candidates\\{\scriptsize (passed vetting)}};

\node[decision] (gaia) at (3.2,0) {Sources in 30\arcsec\\$G < G_{\max}$?};

\node[decision] (ruwe1) at (6.2,1.3) {RUWE\\$<1.4$?};
\node[decision] (ruwe2) at (6.2,-1.3) {RUWE\\$<1.4$?};

\node[validated] (pc) at (9.2,2.0) {PLANET\\CANDIDATE};
\node[caution] (heb) at (9.2,0.6) {PC\_HEB\\CAUTION};
\node[pending] (hr) at (9.2,-0.6) {NEEDS\_HR\\IMAGING};
\node[rejected] (fp) at (9.2,-2.0) {FALSE\\POSITIVE};

\draw[arrow] (input) -- (gaia);
\draw[arrow] (gaia) -- node[yeslabel, above, pos=0.3] {0} (ruwe1);
\draw[arrow] (gaia) -- node[nolabel, below, pos=0.3] {$\geq$1} (ruwe2);
\draw[arrow] (ruwe1) -- node[yeslabel, above, pos=0.4] {yes} (pc);
\draw[arrow] (ruwe1) -- node[nolabel, below, pos=0.4] {no} (heb);
\draw[arrow] (ruwe2) -- node[yeslabel, above, pos=0.4] {yes} (hr);
\draw[arrow] (ruwe2) -- node[nolabel, below, pos=0.4] {no} (fp);

\end{tikzpicture}%
}
\caption{Gaia~DR3 physical verification decision tree. Candidates passing statistical vetting are classified based on aperture contamination ($G_{\rm max}$ from Eq.~\ref{eq:gmax}) and astrometric quality (RUWE).}
\label{fig:decision_tree}
\end{figure}

\subsection{Signal reliability framework}
\label{sec:tier}

The SDE~$\geq 7$ detection threshold has a $\sim$17\% empirical false alarm rate on M3-M6 dwarfs (Sect.~\ref{sec:far}), motivating a per-candidate reliability assessment beyond the binary detected or not-detected classifications. Three independent empirical tests probe complementary aspects of signal robustness:

\begin{enumerate}
    \item Circular shift FAR \citep{Jenkins2002}: per-sector time shifts destroy inter-sector transit coherence while preserving stellar variability. A TIC is flagged if the scrambled light curve produces SDE~$\geq 7$ candidates through the full pipeline;
    \item Light-curve inversion \citep{Coughlin2016}: reflecting the flux ($f_{\rm inv}=2 - f$) converts transit dips to bumps; any SDE~$\geq 7$ detection in the inverted data establishes the per-star noise floor. A candidate is flagged if SDE$_{\rm inv} \geq 0.8 \times \mathrm{SDE}_{\rm real}$;
    \item Fourier phase scrambling \citep{Timmer1995}: per-sector FFT with randomized phases preserves the power spectrum but destroys all waveform coherence. The per-star false alarm probability (FAP) is then estimated from $N=20$ scrambles (resolution $\Delta\mathrm{FAP}=5\%$).
\end{enumerate}

\noindent The combined results classify each candidate into one of three signal reliability tiers (Table~\ref{tab:tier_definition}). The astrophysical nature assessment (BEB exclusion, FPP, centroid, odd-even) is reported separately from signal reliability.

\begin{table}[!htbp]
\centering
\caption{Signal reliability tier definitions.}
\label{tab:tier_definition}
\footnotesize
\begin{tabular}{@{}cl@{}}
\toprule
\textbf{Tier} & \textbf{Signal criteria} \\
\midrule
1 (High robustness) & No FAR flag AND SDE$_{\rm inv} < 7$ \\
2 (Moderate robustness) & Flagged by one test only \\
3 (Noise-susceptible) & Flagged by both tests \\
\bottomrule
\end{tabular}
\tablefoot{FAR flag=host TIC produced $\geq$1 false alarm candidate on circularly shifted light curve. SDE$_{\rm inv}$=best SDE from inverted light curve (noise floor). Fourier FAP refines the initial classification: FAP~$\geq 25\%$ downgrades one tier. Monotransit candidates (single transit epoch) are excluded from the tier system. Detailed results and per-candidate tier assignments are presented in Sect.~\ref{sec:far}.}
\end{table}

\section{Injection-recovery tests}
\label{sec:injection}

The pipeline detection sensitivity is characterized through end-to-end injection-recovery tests on activity-stratified representative hosts. The 121-star sample spans a wide range of photometric scatter ($\sigma_{\rm MAD}=1\,204$--$10\,801$~ppm), and detection sensitivity varies strongly with stellar activity. To capture this dependence, three archetype hosts were selected from the tercile boundaries of the sample scatter distribution: a photometrically calm star (TIC~402375302; $\sigma_{\rm MAD}=1\,519$~ppm), a median-activity star (TIC~186664624; $\sigma_{\rm MAD}=4\,559$~ppm), and an active star (TIC~129781877; $\sigma_{\rm MAD}=5\,870$~ppm). Each archetype received 100 injections using four-dimensional Sobol quasi-Monte Carlo (QMC) sampling \citep[$R_p$, $P$, $b$, $T_0$;][]{Ment2023}. Unlike occurrence rate studies requiring dense Monte Carlo sampling across the parameter grid, this injection-recovery campaign characterizes detection degradation across activity regimes, adopting the sparse sampling approach of \citet{Ment2023} with $\pm 16\%$ Poisson uncertainty per radius-period bin (ten injections per bin).

\subsection{Methodology}
\label{sec:injection_method}

Synthetic transits were generated with \texttt{batman} \citep{Kreidberg2015} and injected into raw lightcurves before any preprocessing. Each injection was processed through the full pipeline, consisting of GP detrending followed by a TLS detection with a Scout+Sniper optimization (Sect.~\ref{sec:search}), referrred to as subprocesses to ensure identical treatment to the discovery pipeline. The injection parameters were:

\begin{itemize}
    \item Planet radius: $R_p \in [0.5, 4.0]~R_\oplus$, log-uniform;
    \item Orbital period: $P \in [0.5, 100]$~d, log-uniform;
    \item Impact parameter: $b \sim \mathcal{U}(0, 0.9)$; eccentricity $e=0$ (circular orbits);
    \item Epoch: $T_0 \sim \mathcal{U}(t_{\rm min}, t_{\rm min} + P)$;
    \item Limb darkening: quadratic coefficients for the TESS bandpass from \citet{Claret2017}, interpolated per stellar $T_{\mathrm{eff}}$;
    \item Stellar parameters: $R_\star$ and $M_\star$ from the TIC for each target.
\end{itemize}

A signal was considered recovered if $\mathrm{SDE} \geq 7.0$ and $\min_k |P_{\rm found} - P_k|/P_k < 0.002$, where $P_k \in \{nP_{\rm inj},\, P_{\rm inj}/n\}$ for $n \in \{1, 2, \ldots, 7\}$, following \citet{Ment2023}. The 0.2\% tolerance ensures precise period matching; the harmonic set accounts for TLS phase-folding degeneracies at integer multiples of the true period \citep{Hippke2019}. Fractional-ratio aliases (e.g., $2P/3$, $4P/5$) are excluded for consistency with the literature; their inclusion would increase recovery rates by $\sim$2 percentage points.

The period range was extended to $P=100$~d, covering the full accessible detection space including extended periods ($P=20$--$100$~d) where sparse coverage limits detection. These tests measure detection completeness only. The vetting stage (TRICERATOPS, centroid analysis) is excluded because it requires pixel-level data unavailable for synthetic injections; vetting produced zero false negatives across 16 validated planets (Sect.~\ref{sec:validation}).

\subsection{Activity-stratified completeness}
\label{sec:global_completeness}

Table~\ref{tab:completeness} summarizes recovery rates for three activity archetypes. The calm archetype achieves $56.0\%$ overall recovery (56/100; 95\% Wilson CI $[46, 65]\%$), the median-activity archetype recovers $28.0\%$ (28/100; CI $[20, 37]\%$), and the active archetype recovers only $18.0\%$ (18/100; CI $[12, 27]\%$)---a total degradation factor of $3.1\times$ from calm to active, consistent with the $\beta \approx 1.5$--$1.6$ reported by \citet{Yaptangco2025} for active versus inactive M-dwarfs.

\begin{table}[!htbp]
\centering
\caption{Activity-stratified injection-recovery completeness.}
\label{tab:completeness}
\scriptsize
\begin{tabular}{lccc}
\toprule
 & Calm$^a$ & Median$^b$ & Active$^c$ \\
\midrule
\textbf{Overall} & \textbf{56/100=56.0\%} & \textbf{28/100=28.0\%} & \textbf{18/100=18.0\%} \\
 & $[46, 65]\%$ & $[20, 37]\%$ & $[12, 27]\%$ \\
\midrule
\multicolumn{4}{l}{\emph{By planet radius}} \\
$R_p < 1.0~R_\oplus$      &  4/33=12.1\% &  0/35=0.0\% &  0/34=0.0\% \\
$1.0 \leq R_p < 2.0~R_\oplus$ & 23/33=69.7\% &  7/31=22.6\% &  4/34=11.8\% \\
$R_p \geq 2.0~R_\oplus$   & 29/34=85.3\% & 21/34=61.8\% & 14/32=43.8\% \\
\midrule
\multicolumn{4}{l}{\emph{By orbital period}} \\
$P < 5$\,d    & 32/44=72.7\% & 20/43=46.5\% & 13/43=30.2\% \\
$5 \leq P < 20$\,d & 15/25=60.0\% &  6/28=21.4\% &  5/27=18.5\% \\
$P \geq 20$\,d & 9/31=29.0\% &  2/29=6.9\% &  0/30=0.0\% \\
\midrule
Min.\ recovered $R_p$ & $0.77~R_\oplus$ & $1.34~R_\oplus$ & $1.76~R_\oplus$ \\
SDE~$\geq 7$, wrong $P$ & 3 & 32 & 0 \\
\bottomrule
\end{tabular}
\tablefoot{100 Sobol QMC injections per host. Wilson 95\% CIs in brackets. $^a$TIC~402375302, $\sigma_{\rm MAD}=1\,519$~ppm. $^b$TIC~186664624, $\sigma_{\rm MAD}=4\,559$~ppm. $^c$TIC~129781877, $\sigma_{\rm MAD}=5\,870$~ppm. Wrong $P$: SDE~$\geq 7$ candidate but no harmonic alias match within 0.2\%.}
\end{table}

Recovery depends strongly on both the planet size and orbital period. For the calm archetype, planets with $R_p \geq 2.0~R_\oplus$ are recovered at 85\% (29/34) across all periods, and short-period planets ($P < 5$~d) at 73\% (32/44). Sub-Earths ($R_p < 1.0~R_\oplus$) are recovered at 12\% (4/33), with the smallest recovered planet at $0.77~R_\oplus$. At extended periods ($P \geq 20$~d), completeness drops to 29\% (9/31) even for the calm archetype; three injections produced wrong-period detections (SDE~$\geq 7$ at non-matching periods).

For the median archetype, no sub-Earth was recovered (0/35), the minimum recovered radius rises to $1.34~R_\oplus$, and long-period recovery ($P \geq 20$~d) drops to 7\% (2/29). Stellar variability is the dominant factor: 32 of 100 injections produced $\mathrm{SDE} \geq 7$ candidates at incorrect periods, with 19 of these locking onto the stellar rotation signal at $P \approx 0.54$~d. This ``variability confusion'' mechanism, where TLS preferentially detects the highest-amplitude periodic signal, explains the discrepancy between the 100\% pipeline validation on known systems (Sect.~\ref{sec:validation}), which were pre-selected as planet hosts, and the lower injection recovery rates on representative M3-M6 targets.

The active archetype ($\sigma_{\rm MAD}=5\,870$~ppm) confirms these trends at the extreme end: overall recovery drops to 18\% (18/100; CI $[12, 27]\%$), no planet below $1.76~R_\oplus$ was recovered and no recovery occurred beyond $P=14.8$~d. Although 65 of 100 injections yielded maximum TLS $\mathrm{SDE} \geq 7$, none of these false-period detections survived the vetting checks (0 wrong-$P$ candidates). The full calm-median-active sequence yields a $3.1\times$ degradation factor at approximately constant $R_\star$ ($0.328$--$0.340~R_\odot$).

The calm archetype achieved $\geq$50\% recovery for $R_p \geq 1.0~R_\oplus$ at $P < 10$~d and for $R_p \geq 2.0~R_\oplus$ at all periods, while the active archetype has been confined to $R_p \geq 2.0~R_\oplus$ at $P < 5$~d. Per-bin recovery fractions (Table~\ref{tab:completeness}) are based on 25--44 injections per bin; the corresponding Wilson 95\% confidence intervals are $\pm 10$--$16\%$ (absolute), comparable to the $\sim$100 injections per archetype used by \citet{Ment2023}. These uncertainties are dominated by the Sobol sampling density (rather than by pipeline stochasticity) and do not affect the main conclusions (calm-to-active degradation, sub-Earth detection threshold), which are derived from the full 100-injection aggregates with tighter constraints (Table~\ref{tab:completeness}, top row).

\section{Pipeline robustness}
\label{sec:validation}

Injection tests characterize the sensitivity to synthetic signals; validation on real planetary systems provides a complementary robustness check. The full detection and vetting chain was applied blindly to 10 M-dwarf systems with confirmed transiting planets that meet the sample selection criteria ($\geq$3 TESS sectors): L~98-59 \citep{Kostov2019} (3 planets), TOI-700 \citep{Gilbert2020} (4 planets including HZ planet d), Gliese~12 \citep{Dholakia2024} (HZ planet b), GJ~357 \citep{Luque2019}, GJ~3473 \citep{Kemmer2020}, GJ~486 \citep{Trifonov2021}, TOI-406, TOI-782, TOI-6086 \citep{Barkaoui2024}, and LP~791-18 \citep{Crossfield2019} (2 TESS-detectable planets). This test achieved 100\% recovery (16/16 TESS-detectable planets) with zero false positives (Table~\ref{tab:validation}). Detailed per-system analysis is provided in Appendix~\ref{app:validation}.

\begin{table}[!htbp]
\centering
\caption{Validation summary: 10 systems, 16 TESS-detectable planets, 100\% recovery.}
\label{tab:validation}
\footnotesize
\begin{tabular}{@{}lccccl@{}}
\toprule
System & SpT & Pl. & HZ & Gaia & Status \\
\midrule
L~98-59 & M3V & 3/3 & --- & 0 & PLANET\_CAND \\
TOI-700 & M2V & 4/4 & d & 2--5 & NEEDS\_HR \\
Gliese~12 & M4V & 1/1 & b & 0 & PLANET\_CAND \\
GJ~357 & M2.5V & 1/1 & --- & 2 & NEEDS\_HR \\
GJ~3473 & M4V & 1/1 & --- & 0 & PLANET\_CAND \\
GJ~486 & M3.5V & 1/1 & --- & 0 & PLANET\_CAND \\
TOI-406 & M4V & 1/1 & --- & 1 & NEEDS\_HR \\
TOI-782 & M3V & 1/1 & --- & 1 & NEEDS\_HR \\
TOI-6086 & M3.5V & 1/1 & --- & 1 & NEEDS\_HR \\
LP~791-18 & M5V & 2/2$^\dagger$ & --- & 1--2 & NEEDS\_HR \\
\midrule
\textbf{Total} & & \textbf{16/16} & \textbf{2} & & \textbf{0 FP} \\
\bottomrule
\end{tabular}
\tablefoot{Gaia=contaminants within 30$''$. $^\dagger$LP~791-18~d excluded (Spitzer discovery, \citealt{Peterson2023}). All validation planets classified as NEEDS\_HR\_IMAGING were confirmed using speckle or AO imaging in their discovery papers \citep{Gilbert2020, Luque2019, Barkaoui2024, Peterson2023}.}
\end{table}

The validation demonstrates several important results. Period accuracy exceeds 99.99\% for 15 of 16 detections; L~98-59~d shows a 0.94\% offset attributable to iterative planet masking (Table~\ref{tab:app_val_detection}). Both HZ validation planets (TOI-700~d and Gliese~12~b) were correctly identified. The \texttt{NEEDS\_HR\_IMAGING} classification is the majority outcome (10 of 16 planets), consistent with the requirements from the original discovery papers. Validation targets satisfy the sector criterion (at least three sectors) but not necessarily the M3--M6 spectral type restriction (e.g., TOI-700 is M2V). No M6 validation targets meeting the sample criteria exist. LP~791-18 (M5V, $T_{\mathrm{eff}}=2949$~K) validates the detection on late M-dwarfs, although its magnitude ($T_{\rm mag}=13.56$) exceeds the sample limit. Candidates around M6 hosts remain less robustly validated than earlier types. The signal reliability tier system (Sect.~\ref{sec:tier}) was not formally applied to validation targets: these hosts have photometric scatter well below the noise frontier (median MAD~$\approx 1\,800$~ppm versus $4\,800$~ppm for the survey sample) and detection SDE~$= 18$-$65$ (versus $7$-$12$ for survey candidates). Thus, the tier system (designed to discriminate at the noise frontier) would trivially assign tier~1 to all validation hosts, providing no calibration information at the SDE and noise levels where the classification appears ambiguous.

\begin{figure}[!htbp]
\centering
\includegraphics[width=0.49\columnwidth,height=2.65cm]{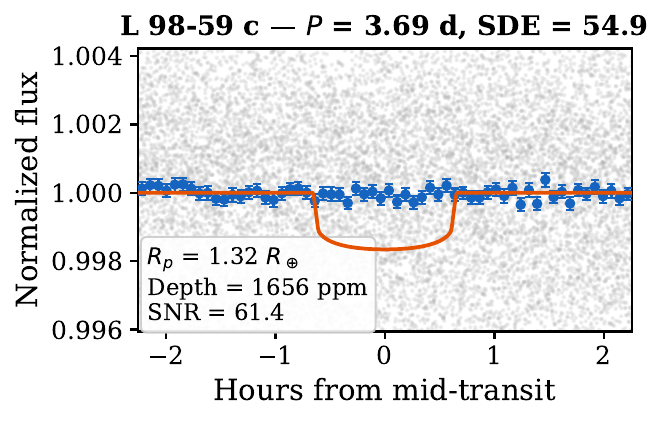}\hfill
\includegraphics[width=0.49\columnwidth,height=2.65cm]{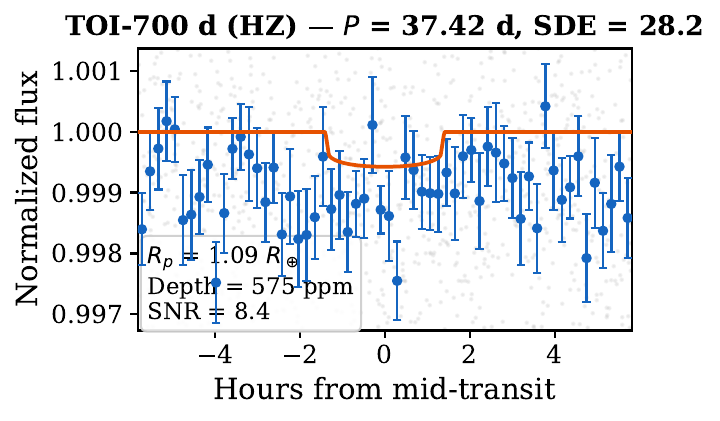}\\[0.2em]
\includegraphics[width=0.49\columnwidth,height=2.65cm]{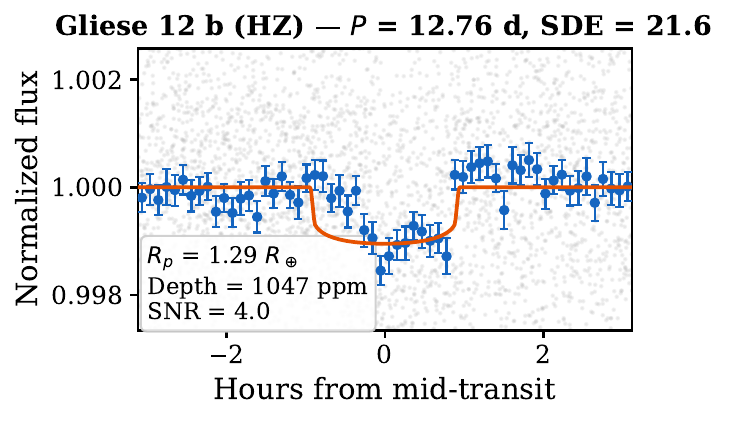}\hfill
\includegraphics[width=0.49\columnwidth,height=2.65cm]{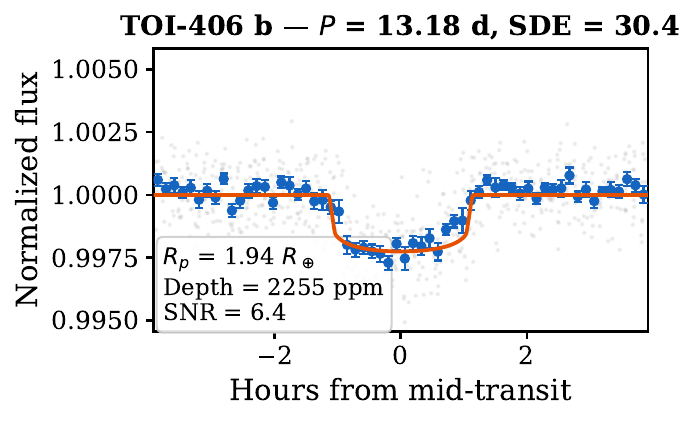}
\caption{Validation light curves (1/2). Top row: L~98-59~c, TOI-700~d (HZ). Bottom row: Gliese~12~b (HZ), TOI-406~b.}
\label{fig:val_1}
\end{figure}

\begin{figure}[!htbp]
\centering
\includegraphics[width=0.49\columnwidth,height=2.65cm]{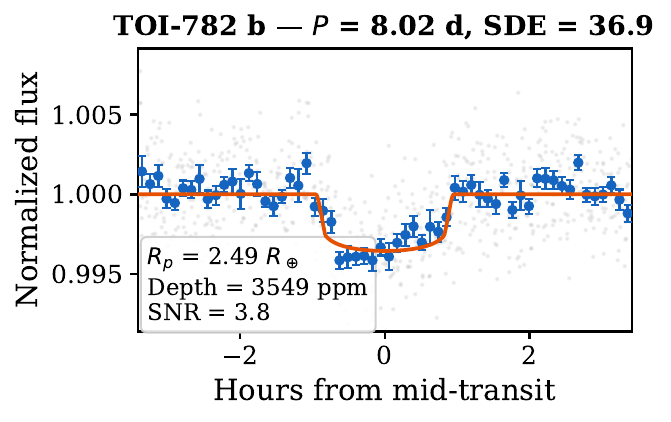}\hfill
\includegraphics[width=0.49\columnwidth,height=2.65cm]{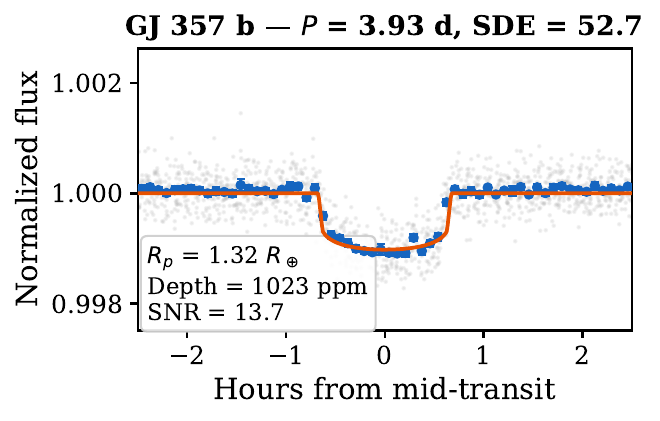}\\[0.2em]
\includegraphics[width=0.49\columnwidth,height=2.65cm]{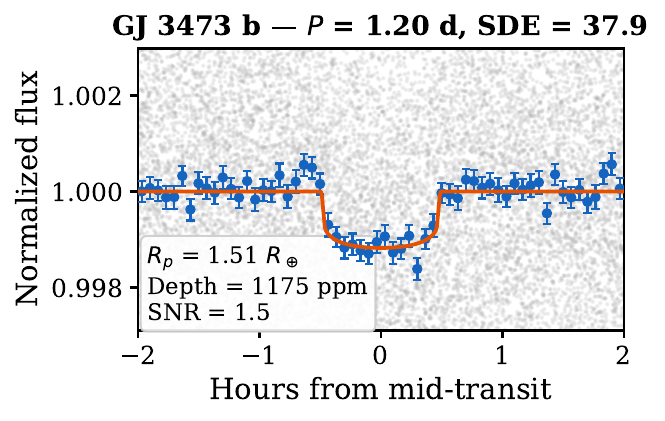}\hfill
\includegraphics[width=0.49\columnwidth,height=2.65cm]{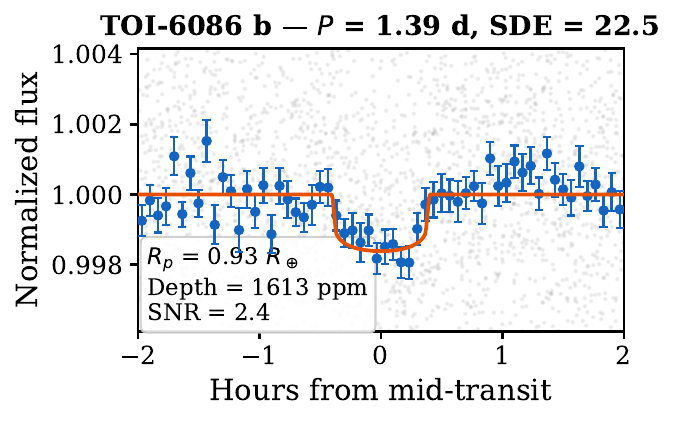}
\caption{Validation lightcurves (2/2). Top row: TOI-782~b, GJ~357~b. Bottom row: GJ~3473~b, TOI-6086~b.}
\label{fig:val_2}
\end{figure}

Phase-folded light curves are shown in Figs.~\ref{fig:val_1} and \ref{fig:val_2}.

\FloatBarrier
\section{Limitations and failure modes}
\label{sec:limitations}

Characterizing pipeline limitations is essential for interpreting detection statistics and occurrence rate estimates. The sample selection (Table~\ref{tab:funnel}) was designed to mitigate several known failure modes, but residual limitations remain.

\paragraph{Photometric quality.} Ultra-bright stars ($T_{\rm mag} < 8$) exhibit severe red noise rendering transit detection unreliable; testing on AU~Mic \citep[$T_{\rm mag}=6.8$;][]{Plavchan2020} yielded 0/3 planet recoveries in spite of the availability of three sectors and deep transits (${\sim}2\,000$~ppm), with a red noise factor $\beta_{\rm rn}=10.0$ (the survey maximum). Ultra-faint stars ($T_{\rm mag} > 13$) suffer photon-limited noise: testing on LP~890-9 ($T_{\rm mag}=14.3$, 4~sectors) recovered planet~b at the correct period ($P=2.73$~d, SDE~$= 18.5$) but missed the HZ planet~c ($P=8.46$~d), whose discovery required 614~hours of ground-based SPECULOOS observations \citep{Delrez2022}. Both failure modes are mitigated by sample selection ($T_{\rm mag}=8$--$12.5$), though LP~890-9 demonstrates that short-period planets with deep transits remain detectable beyond this range.

\paragraph{Period blind spots.} Periods near TESS systematics, namely, 0.5~d, 1.0~d (Earth rotation; \citealt{Fausnaugh2018}), and 13.7~d (spacecraft orbit), were rejected by the vetting cascade (Sect.~\ref{sec:signal_validation}), creating blind spots where genuine ultra-short-period planets would be missed. The harmonic breaker (Sect.~\ref{sec:subharmonic}) corrects subharmonic aliases ($P/2$, $P/3$) but requires sufficient phase coverage to distinguish the true period; with sparse coverage ($\leq$5 sectors), this correction can fail (e.g., L~98-59~c detected at $P/2$ under three-sector conditions).

\paragraph{Scout+Sniper limitations.} The two-phase optimization introduces three potential failure modes: (1) Scout threshold losses (signals with $4.5 < \mathrm{SDE}_{\rm binned} < 7$ may be missed by Scout); meanwhile (2) Sniper window misses for strong aliases outside $\pm 10\%$ of the Scout period; (3) ultra-short transit dilution for $P < 1$~d around $R_\star < 0.15~R_\odot$ hosts where transit duration approaches the binning cadence. These losses are included in the injection-recovery completeness (Sect.~\ref{sec:injection}), which provides an end-to-end measure of detection sensitivity including Scout+Sniper effects.

\paragraph{Stellar activity.} \label{sec:activity_limitation} Active M-dwarfs with elevated red noise can push marginal detections below threshold. The AU~Mic limit test demonstrates this: despite having three sectors and deep transits ($\sim$2000~ppm), all known planets were missed due to extreme stellar variability. The pipeline deliberately prioritizes sample purity over completeness on active stars. This trade-off reduces the risk of false positives at the cost of missing some genuine planets.

\paragraph{Sparse temporal coverage.} This is the dominant limitation. The survey targets have between three and ten sectors (with a median of about four), providing only $\sim$80--160~d of photometry spread over baselines of 700--2\,600~d. This sparse coverage has two consequences: (1)~planets with $P > 20$~d produce $\leq$3 transit events, insufficient for robust phase-folding and period determination; (2)~the gap-dominated window function generates period aliases. Validation tests quantify the impact: restricting four known systems (L~98-59, TOI-700, Gliese~12, TOI-406) to 3~sectors recovers only 2/9 planets (22\%), with all five planets at $P > 10$~d missed regardless of depth. With full sector coverage, the same systems achieve 9/9 (100\%). The detection capability therefore scales directly with sector count and as the ``newly enabled'' targets accumulate additional TESS sectors, completeness at longer periods will improve.

\section{Results}
\label{sec:results}

\subsection{Survey statistics}

Table~\ref{tab:survey_stats} summarizes the detection statistics from the complete survey.

\begin{table}[!htbp]
\centering
\caption{Survey detection statistics for the 121-star M3--M6 sample.}
\label{tab:survey_stats}
\begin{tabular}{lcc}
\toprule
\textbf{Category} & \textbf{Count} & \textbf{Fraction} \\
\midrule
Total stars analyzed & 121 & 100\% \\
No detection (SDE $< 7$) & 98 & 81.0\% \\
With transit signals (SDE $\geq 7$) & 23 & 19.0\% \\
\midrule
\textbf{Surviving candidates} & \textbf{20} & \textbf{16.5\%}$^a$ \\
\bottomrule
\end{tabular}
\tablefoot{$^a$Fraction of total sample (121 stars).}
\end{table}

Transit-like signals (SDE~$\geq 7$) were detected in 19.0\% of targets (23/121), producing 33 candidates after harmonic nest cleanup. After vetting, 13 candidates were rejected (9 \texttt{LIKELY\_EB}, 3 \texttt{FALSE\_POSITIVE}, 1 \texttt{GRAZING\_CANDIDATE}), yielding 20 surviving transit-like signals across 16 systems (Table~\ref{tab:classification}). The 39\% rejection rate (13/33) is dominated by eclipsing binary scenarios (9 \texttt{LIKELY\_EB}), consistent with the high BEB incidence expected from TESS's large pixel scale ($21\arcsec$) for faint M-dwarf hosts \citep{Sullivan2015}. All 20 surviving signals received the \texttt{HIGH\_CONFIDENCE} quality flag; no \texttt{MARGINAL\_ACTIVE\_STAR} candidate survived to the final classification. None have prior TOI designations in ExoFOP-TESS or the NASA Exoplanet Archive (query date: 28 February 2026). Exhaustive per-candidate parameters, host star properties, and rejected candidate details are provided in Appendix~\ref{app:catalog}. Numerical values of the signal-validation cascade diagnostics (Table~\ref{tab:checks}) for each of the 20 signals are tabulated in Table~\ref{tab:app_vetting_numerics}.

\subsection{Candidate classification}

The 20 signals are classified along two complementary axes: physical verification (\texttt{PLANET\_CANDIDATE} or \texttt{NEEDS\_HR\_IMAGING}, Sect.~\ref{sec:physical_verification}) determines whether astrophysical false positives are excluded; signal reliability (tier~1--3, Sect.~\ref{sec:tier}) quantifies the risk that stellar noise produces a signal of comparable strength. At the stellar noise frontier, signal reliability is the primary classification axis; physical verification determines imaging priorities. Table~\ref{tab:classification} presents the disposition of all detected signals after TRICERATOPS vetting and Gaia DR3 physical verification.

\begin{table}[!htbp]
\centering
\caption{Final classification of detected transit signals.}
\label{tab:classification}
\small
\begin{tabular}{lcc}
\toprule
\textbf{Category} & \textbf{Cand.} & \textbf{Sys.$^a$} \\
\midrule
\multicolumn{3}{l}{\textit{Surviving candidates}} \\
\texttt{PLANET\_CANDIDATE} & 9 & 8 \\
\texttt{NEEDS\_HR\_IMAGING} & 11 & 9 \\
\midrule
\multicolumn{3}{l}{\textit{Rejected candidates}} \\
\texttt{LIKELY\_EB} & 9 & 8 \\
\texttt{FALSE\_POSITIVE} & 3 & 3 \\
\texttt{GRAZING\_CANDIDATE} & 1 & 1 \\
\midrule
\textbf{Total} & \textbf{33} & \textbf{23}$^b$ \\
\bottomrule
\end{tabular}
\tablefoot{Some TICs contribute candidates to multiple categories. $^a$Unique TICs per category. $^b$Total unique TICs with signals; column does not sum due to multicategory TICs (e.g., TIC-24293836 has both PC and HR candidates, plus one LIKELY\_EB).}
\end{table}

\subsection{Signal reliability results}
\label{sec:far}

Individual transit events for the 20 signals have per-epoch S/N~$<1$; the signal emerges only after phase-folding between three and six transits. At SDE~$=7$, the white-noise FAP is $\sim$1\% per TLS test \citep{Hippke2019}; correlated stellar variability increases the effective FAP, which cannot be estimated analytically for BY~Dra lightcurves. The three empirical tests described in Sect.~\ref{sec:tier} were applied to all candidate hosts.

\subsubsection{Circular shift FAR}

A per-sector circular shift scramble test was performed on all 121 targets \citep{Jenkins2002}: each continuous sector, identified by time gaps $>5$~d, receives an independent circular shift (10--90\% of its length), preserving intra-sector autocorrelation while destroying inter-sector transit phase coherence. The full pipeline (TLS detection + 18-check vetting cascade) was then applied to the 121 scrambled lightcurves under identical conditions.

Of the 121 scrambled lightcurves, 21 (17.4\%; Wilson 95\% CI: [11.6\%, 25.1\%]) produced at least one candidate that passed the full search cascade (SDE~$\geq 7$ plus quality checks), yielding 28 individual false candidates. The false-alarm hosts span the full range of sample variability (MAD~$= 1\,600$--$9\,600$~ppm) and their false-alarm rate ($\sim$16\%) does not differ significantly from the base rate when stratified by photometric scatter, indicating that after GP detrending the residual noise level is sufficient to generate spurious SDE~$\geq 7$ detections regardless of raw stellar activity. The false-alarm SDE distribution clusters near the detection threshold: median~$= 8.9$, interquartile range $7.8$--$9.8$, maximum~$= 14.1$. This maximum underscores that SDE alone cannot distinguish real from spurious signals on M-dwarfs with sparse multisector coverage \citep[see also][]{Pont2006}.

\subsubsection{Light-curve inversion test}

A light-curve inversion procedure \citep{Coughlin2016} reflects the flux around the baseline ($f_{\rm inv}=2 - f$), converting any real transit dip into a bump while preserving the noise variance. Any $\mathrm{SDE}_{\rm inv} \geq 7$ detection in the inverted data therefore establishes a per-star noise floor that reaches the TLS detection threshold. Across the full 121-target sample, 97 hosts (80.2\%) have $\mathrm{SDE}_{\rm inv} < 7$: their noise floors lie below the TLS detection threshold and remain unresolved by this test. The remaining 24 hosts (19.8\%) reach $\mathrm{SDE}_{\rm inv}=7$--$12.4$, providing a measurable per-star floor; the whole-sample distribution and its correlation with $\sigma_{\rm MAD}$ are presented in Appendix~\ref{app:inversion_whole}. Among the 16 candidate hosts, 7 belong to the BELOW group ($\mathrm{SDE}_{\rm inv} < 7$, candidate signals sit above an unresolved floor), 5 are noise-comparable (COMP, $0.8 \leq \mathrm{SDE}_{\rm inv}/\mathrm{SDE}_{\rm real} < 1$), and 4 are noise-dominated (DOM, $\mathrm{SDE}_{\rm inv} \geq \mathrm{SDE}_{\rm real}$); for these 9 measurable cases ($\mathrm{REAL\_EXCEEDS\_NOISE}=0$) the detections are not cleanly separable from stellar noise at the current sensitivity. This split is a direct consequence of searching for small planets (100--2\,000~ppm depth) on active M3--M6 dwarfs (median scatter of 4\,800~ppm).

\subsubsection{Fourier per-star FAP}

To quantify per-candidate significance, Fourier phase scrambling \citep{Timmer1995} was applied to all 16 candidate host TICs: per-sector FFT with randomized phases destroys all waveform coherence, while preserving the power spectrum. For each TIC, $N=20$ scrambles were processed through the full pipeline (resolution $\Delta\mathrm{FAP}=5\%$). The FAP is the fraction of scrambles producing SDE~$\geq$ SDE$_{\rm real}$.

Of the 16 TICs, 8 (50\%) show FAP~$= 0\%$ (0/20 exceedances; 95\% CI~$< 16.8\%$ by the rule of three), two are marginal (FAP~$= 10$--$20\%$), and six are noise-limited (FAP~$\geq 25\%$). The two highest-MAD hosts ($\sigma_{\rm MAD} > 7\,500$~ppm) have the worst FAP ($\geq 45\%$), consistent with elevated stellar noise driving both real and spurious SDE peaks. Notably, three candidates were dual-flagged by FAR and inversion but still achieved FAP~$= 0\%$, indicating that their waveform morphology is not reproducible by noise alone$-$even though the host star's noise floor reaches comparable SDE; these candidates warrant priority monitoring for signal persistence. Per-TIC FAP values are reported in Table~\ref{tab:tier_classification}. A FAP~$\geq 25\%$ downgrades the assignment by one tier (Sect.~\ref{sec:tier}).

\subsubsection{Combined tier classification}

Crossing the FAR results (per-TIC flag) with the inversion results (per-TIC noise floor) yields an initial tier assignment: six candidates clean on both tests, eight single-flagged, and six dual-flagged. Fourier FAP refinement (Sect.~\ref{sec:tier}) then modifies four systems: TIC~117955697 (FAP~$= 25\%$, tier~1$\to$2), TIC~77634567 (FAP~$= 45\%$, tier~1$\to$2), TIC~186664624 (FAP~$= 40\%$, tier~2$\to$3), and TIC~24293836 (FAP~$= 25\%$, tier~2$\to$3). TIC~72750190 is excluded from the tier system as a monotransit candidate \citep[single transit epoch, MCMC unconverged;][]{Osborn2016, Cooke2021}.

\begin{figure}[!htbp]
\centering
\includegraphics[width=0.49\columnwidth,height=2.65cm]{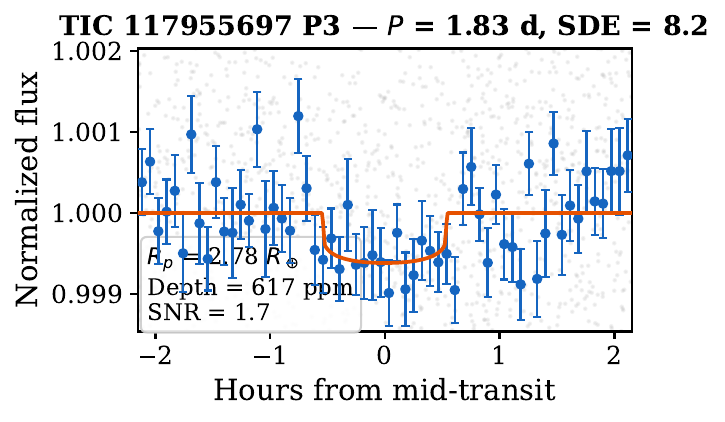}\hfill
\includegraphics[width=0.49\columnwidth,height=2.65cm]{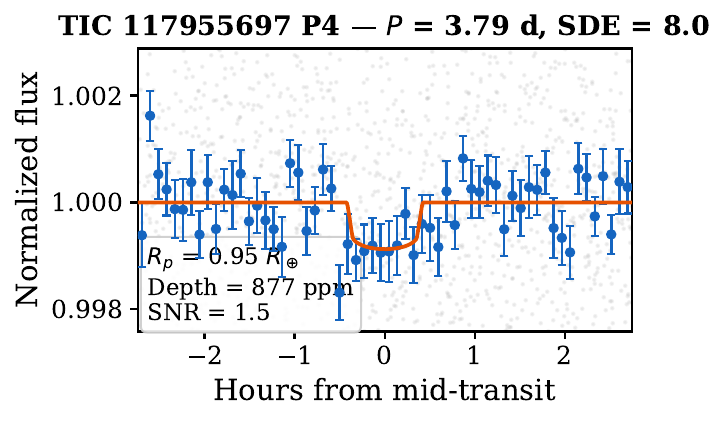}\\[2pt]
\includegraphics[width=0.49\columnwidth,height=2.65cm]{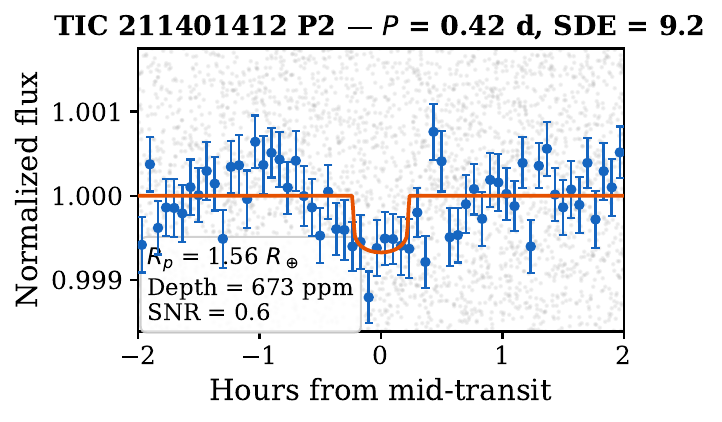}\hfill
\includegraphics[width=0.49\columnwidth,height=2.65cm]{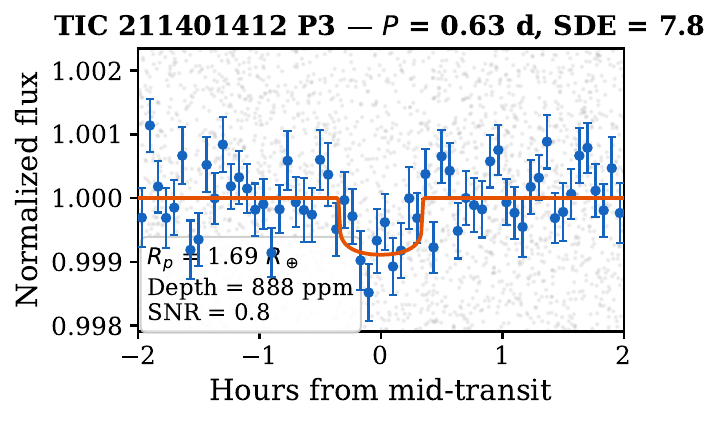}\\[2pt]
\includegraphics[width=0.49\columnwidth,height=2.65cm]{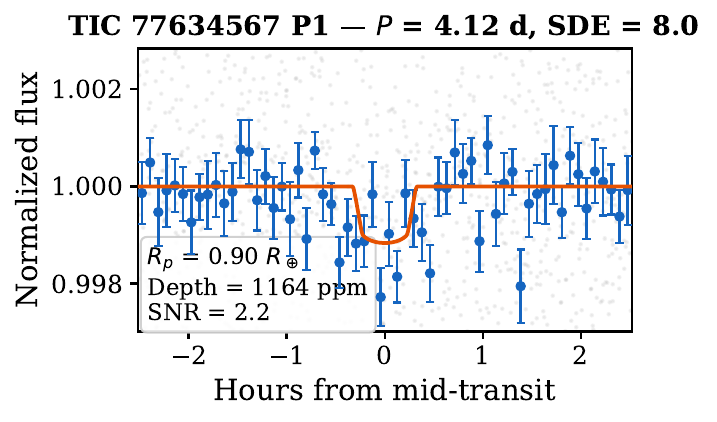}\hfill
\includegraphics[width=0.49\columnwidth,height=2.65cm]{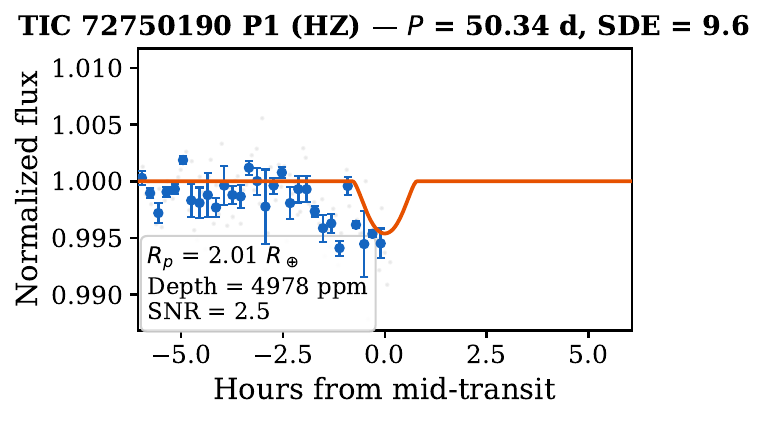}
\caption{Phase-folded lightcurves for the two tier~1 candidates (TIC~211401412), two tier~2 candidates downgraded by Fourier FAP (TIC~117955697, TIC~77634567), and the monotransit candidate TIC~72750190. Top row: TIC~117955697~P1 ($P=1.83$~d, tier~2) and P2 ($P=3.79$~d, tier~2). Middle row: TIC~211401412~P1 ($P=0.42$~d, tier~1) and P2 ($P=0.63$~d, tier~1). Bottom row: TIC~77634567 ($P=4.12$~d, tier~2) and TIC~72750190 ($P=50.3$~d, monotransit, tentative HZ). Grey points: individual TLS phase-folded measurements. Blue circles with error bars: mean-binned data (60 bins, SEM). Orange curve: \texttt{batman} transit model \citep{Kreidberg2015} with quadratic limb darkening (TESS bandpass; \citealt{Claret2017}); shape parameters from MCMC posterior medians, depth scaled to match the TLS-observed transit depth. TIC~72750190 uses a TLS box-model estimate (MCMC did not converge; single transit). S/N values displayed in each panel are per-transit; the phase-folded (combined) S/N equals $\sqrt{N_{\rm tr}/\beta_{\rm rn}}$ times the per-transit value, where $N_{\rm tr}$ is the number of observed transits (Table~\ref{tab:app_vetting_numerics}) and $\beta_{\rm rn}$ is the Pont+ 2006 red-noise variance ratio defined in Sect.~\ref{sec:search}.}
\label{fig:phasefold_tier1}
\end{figure}

The final classification of the 19 tier-assessed candidates is two tier~1 (high robustness), seven tier~2 (moderate robustness), and ten tier~3 (noise-susceptible), with 1 additional monotransit candidate (Table~\ref{tab:tier_classification}). Phase-folded lightcurves for the two tier~1 candidates and selected tier~2 candidates are shown in Fig.~\ref{fig:phasefold_tier1}.

The ``PLANET\_CANDIDATE'' designation reflects this statistical context: these are transit-like signals warranting follow-up and they are not confirmed planets. The two tier~1 candidates (both from TIC~211401412) are the highest-priority targets for RV confirmation. The seven tier~2 candidates require additional photometric confirmation (continued TESS monitoring) or RV pre-screening before resource-intensive follow-up. The ten tier~3 candidates cannot be distinguished from stellar noise at the current photometric sensitivity; thus, additional TESS sectors are needed to establish signal persistence before any follow-up investment is justified.

\begin{table}[!htbp]
\centering
\caption{Combined signal reliability and physical verification classification for all 20 signals.}
\label{tab:tier_classification}
\scriptsize
\begin{tabular}{@{}lccccccc@{}}
\toprule
\textbf{TIC} & \textbf{P\#} & \textbf{$R_p$} & \textbf{FAR} & \textbf{Inv.} & \textbf{FAP} & \textbf{Tier} & \textbf{Status} \\
\midrule
\multicolumn{8}{l}{\textit{Tier 1$-$High robustness (2)}} \\
211401412 & 1 & 1.56 & Clean & BELOW & 0\% & 1 & HR \\
211401412 & 2 & 1.69 & Clean & BELOW & 0\% & 1 & HR \\
\midrule
\multicolumn{8}{l}{\textit{Tier 2$-$Moderate robustness (7)}} \\
417927999 & 1 & 0.76 & FAR & BELOW & 0\% & 2 & PC \\
117955697 & 1 & 2.78 & Clean & BELOW & 25\% & 2$^\dagger$ & PC \\
117955697 & 2 & 3.12 & Clean & BELOW & 25\% & 2$^\dagger$ & PC \\
77634567 & 1 & 0.99 & Clean & BELOW & 45\% & 2$^\dagger$ & PC \\
266523603 & 1 & 0.68 & Clean & COMP & 0\% & 2 & HR \\
124936003 & 1 & 0.69 & Clean & COMP & 0\% & 2 & HR \\
322053216 & 1 & 0.96 & FAR & BELOW & 20\% & 2 & HR \\
\midrule
\multicolumn{8}{l}{\textit{Tier 3$-$Noise-susceptible (10)}} \\
24293836 & 2 & 1.16 & Clean & DOM & 25\% & 3$^\dagger$ & PC \\
129766762 & 1 & 0.84 & FAR & DOM & 30\% & 3 & PC \\
139471702 & 1 & 1.57 & FAR & DOM & 10\% & 3 & PC \\
80315364 & 1 & 0.79 & FAR & DOM & 95\% & 3 & PC \\
24293836 & 1 & 1.72 & Clean & DOM & 25\% & 3$^\dagger$ & HR \\
419939149 & 1 & 1.01 & FAR & COMP & 0\% & 3 & HR \\
123368775 & 1 & 0.71 & FAR & COMP & 0\% & 3 & HR \\
65516102 & 1 & 1.00 & FAR & COMP & 0\% & 3 & HR \\
186664624 & 2 & 2.31 & FAR & BELOW & 40\% & 3$^\dagger$ & HR \\
186664624 & 1 & 1.43 & FAR & BELOW & 40\% & 3$^\dagger$ & HR \\
\midrule
\multicolumn{8}{l}{\textit{Monotransit (1)}} \\
72750190 & 1 & 2.01 & Clean & BELOW & 0\% & M$^a$ & PC \\
\bottomrule
\end{tabular}
\tablefoot{The 19 tier-assessed signals are classified by FAR $\times$ inversion $\times$ Fourier FAP; TIC~72750190 is reported separately as a monotransit candidate. $R_p$ in $R_\oplus$. FAR: Clean=no false alarm on a scrambled light curve; FAR=$\geq$1 false alarm detected. Inv.: BELOW=SDE$_{\rm inv} < 7$; COMP=$0.8 \leq$ SDE$_{\rm inv}$/SDE$_{\rm real} < 1$; DOM=SDE$_{\rm inv} \geq$ SDE$_{\rm real}$. FAP: Fourier per-star FAP ($N=20$ scrambles). $^\dagger$tier modified by FAP refinement (FAP~$\geq 25\%$ downgrades one tier). Status: PC=\texttt{PLANET\_CANDIDATE} (BEB excluded); HR=\texttt{NEEDS\_HR\_IMAGING} (Gaia contaminants present). $^a$Single transit epoch, MCMC unconverged; excluded from tier system \citep[monotransit treatment per][]{Osborn2016, Cooke2021}.}
\end{table}

\section{Discussion}
\label{sec:discussion}

In the detection of 20 transit-like signals, two achieved a high robustness score (tier~1), seven achieved moderate robustness (tier~2), ten are noise-susceptible (tier~3), and one is an unclassified monotransit. These results, based on a targeted 121-star survey at the stellar noise frontier, warrant examination in the context of theoretical expectations, detection limitations, and the prospects for confirmation.

\subsection{Survey strategy in context}

The ``newly enabled'' selection strategy ($\leq$2 archival + $\geq$1 recent sector) deliberately targets a population overlooked by existing surveys: stars where multisector transit detection became possible only with TESS Cycle~6+. This complements rather than competes with SPOC/QLP real-time processing (limited to $P < 13$~d), QLP multisector searches ($T_{\rm mag} < 10.5$), and archival continuous viewing zone (CVZ) surveys focusing on well-observed targets ($>$13 sectors). The survey's primary strength is its sensitivity to small planets around the smallest stars: the M3--M6 restriction ensures transit depths $>$800~ppm for Earth-sized planets, enabling the sub-Earth detections that constitute the most distinctive aspect of the catalog.

A consequence of the ``newly enabled'' selection is that temporal coverage is inherently sparse: typically between three and ten sectors (with a median of four) spread across two to four TESS cycles, producing heavily gapped baselines (700--2\,600~d of baseline with only $\sim$80--160~d of photometry). This creates severe window-function artifacts, particularly at long periods, where three or fewer transit events are observed. The three-sector limit test confirms that 0/5 planets with $P > 10$~d are recovered at the survey minimum, and harmonic aliasing is exacerbated (e.g., L~98-59~c detected at $P/2$). The 18-check signal validation cascade (Sect.~\ref{sec:signal_validation}) and the evidence-based harmonic breaker (Sect.~\ref{sec:subharmonic}) mitigate these effects, but cannot entirely eliminate them. The 81\% non-detection rate (98/121) is expected given the combination of sparse temporal sampling and BY~Dra stellar variability on active M-dwarfs.

\subsection{Detection fraction and geometric expectations}

The 16/121 $= 13.2\%$ candidate-host fraction exceeds the geometric transit expectation by a factor of ${\sim}1.3$. For M3--M6 hosts ($R_\star \approx 0.25~R_\odot$, $M_\star \approx 0.25~M_\odot$), the per-planet transit probability $p_{\rm tr}=R_\star/a$ ranges from $12.5\%$ at $P=0.5$~d to $0.43\%$ at $P=100$~d, with a log-uniform mean of $\langle p_{\rm tr} \rangle=4.1\%$ over the survey period range. Combined with the small-planet occurrence rate of 2.5 per M-dwarf \citep{Dressing2015}, the expected fraction of stars with at least one transiting planet is $1 - (1 - 0.041)^{2.5}=10.0\%$, even under the assumption of near-co-planar systems. This host-level expression accounts for multiplanetary systems by construction: a host is counted once regardless of how many of its planets transit, matching the observed quantity. The excess attributes to the ten tier~3 (noise-susceptible) signals across eight distinct host systems, whose SDEs cannot be distinguished from stellar-noise-induced peaks. After removing these tier~3-only hosts, the remaining eight hosts give $6.6\%$---below the geometric expectation, consistent with the completeness loss on active stars documented in Sect.~\ref{sec:activity_limitation}. 

\subsection{Methodology transferability and testable predictions}

The three key contributions of this work are: the 18-check signal validation cascade, the three-test reliability framework (FAR, inversion, Fourier FAP), and the tier classification system. All three address a general problem in time-domain astrophysics with respect to classifying transit-like signals when correlated stellar noise produces false alarms at rates comparable to genuine detections. This problem extends well beyond the M3-M6 sample studied here.

\paragraph{Other noise-frontier regimes.} The framework transfers without modification to several adjacent populations. K5--M2 dwarfs ($R_\star \approx 0.4$--$0.6~R_\odot$) are entering multisector TESS coverage as the mission accumulates observations and produce an analogous ``newly enabled'' problem with shallower transits (300--800~ppm) and a higher expected tier~3 fraction. Young and pre-main sequence M-dwarfs ($\lesssim$1~Gyr) exhibit BY~Dra modulation 5--50$\times$ larger than field stars \citep{Newton2017}, exacerbating the noise frontier; the FAR and inversion tests provide a per-star noise calibration that analytical false-alarm probabilities cannot capture on non-Gaussian lightcurves. Ultracool dwarfs (M7--L) targeted by ground-based surveys \citep[SPECULOOS, TRAPPIST;][]{Sebastian2021, Gillon2017trappist} have deeper transits ($>$1\%) but compounded red noise from atmospheric and telluric sources; the Fourier phase-scrambling test, which is agnostic to the physical origin of the noise, is particularly relevant here. The faint end of the PLATO sample \citep{Rauer2024} will face an analogous noise-frontier problem driven by correlated instrumental systematics rather than stellar activity, where the same three-test classification layer applies.

\paragraph{Testable prediction: TESS continuous viewing zone.} The principal limitation of this survey is sparse temporal coverage (median of four sectors, $\sim$80~d of photometry), which motivates a concrete, testable prediction. The TESS CVZ provides M3--M6 dwarfs with 30-44 sectors accumulated over six years of observations. The $\sqrt{N}$ scaling of SDE with transit count predicts a $\sim$3.4$\times$ gain, sufficient to lift genuine signals above the noise floor. Three outcomes are predicted: (1)~some tier~1--2 signals persist and strengthen with additional sectors, confirming planetary origin; (2)~all signals remain at the noise floor despite $10\times$ more data, establishing a stellar variability ceiling for TLS detection on M-dwarfs; or (3)~new signals emerge that were undetectable with sparse coverage, constraining the occurrence rate of small planets around active M-dwarfs. Each outcome directly extends the noise frontier characterization established here and informs the optimal resource allocation for follow-up campaigns.

\section{Conclusions}
\label{sec:conclusions}

This survey of 121 ``newly enabled'' M3--M6 dwarfs identified 20 transit-like signals across 16 systems, none of which had a prior TOI designation. The detection parameter space of TLS on active M-dwarfs with sparse multisector coverage was also mapped. The main results are listed below.

\begin{enumerate}
    \item A validated pipeline: 100\% recovery on 16 known planets with 0 false positives. The three-test signal reliability framework (circular shift FAR, light-curve inversion, Fourier phase scrambling) provides a transparent classification system applicable to any noise-dominated detection regime.

    \item Where planets are detectable: injection tests identify the parameter space where the pipeline successfully recovers planets despite stellar noise: short periods ($P < 5$~d, recovery 73\% on calm hosts), large radii ($R_p \geq 2~R_\oplus$, recovery 85\%), and late M-dwarfs (M5--M6) where the $(R_p/R_\star)^2$ depth advantage compensates for elevated variability. Two tier~1 candidates on TIC~211401412 (FAP~$= 0\%$, clean FAR, SDE below noise floor) are the highest-priority targets for RV confirmation.

    \item The noise frontier: heterogeneous across the sample, whereby for 9/16 candidate hosts, the measurable floor equals or exceeds the candidate SDE (REAL\_EXCEEDS\_NOISE~$= 0$), while on 7/16 the floor lies below the TLS detection threshold and the candidates are above their unresolved floor. The global FAR is $17.4\%$ (Wilson 95\% CI: $[11.6\%, 25.1\%]$). With 3--10 sectors, stellar variability on M3--M6 dwarfs produces false detections at rates indistinguishable from transit signals (SDE~$\sim 7$--$12$), confirming a $3.1\times$ recovery degradation from calm to active hosts \citep{Yaptangco2025}.

    \item Pathways to discovery: the TESS continuous viewing zone (30--44 sectors, $\sim$3.4$\times$ SDE gain) can resolve the classification ambiguity for the signals identified here. Beyond TESS, the tier framework and validation cascade are directly transferable to K5--M2 dwarfs entering multisector coverage, young stellar populations where variability amplitudes are 5--50$\times$ larger ($<$1~Gyr; K2 clusters, TESS young groups), ultracool dwarfs targeted by ground-based surveys \citep[SPECULOOS;][]{Sebastian2021}, and the faint end of the PLATO sample \citep{Rauer2024}. In each case, the pipeline maps the boundary between detectable and noise-dominated parameter space, enabling targeted allocation of follow-up resources.
\end{enumerate}

\section*{Data availability}

Tables~\ref{tab:app_pc}, \ref{tab:app_pc_stars}, \ref{tab:app_hr}, \ref{tab:app_hr_facilities}, \ref{tab:app_grazing}, \ref{tab:app_eb}, \ref{tab:app_fp}, \ref{tab:app_vetting_numerics}, \ref{tab:app_val_overview}, and \ref{tab:app_val_detection} are only available in electronic form at the CDS via anonymous ftp to \texttt{cdsarc.u-strasbg.fr} (130.79.128.5) or via \url{http://cdsweb.u-strasbg.fr/cgi-bin/qcat?J/A+A/}. Pipeline code is available at \url{https://gitlab.com/yohanntschudi/exoplanets-project}. TESS photometry was obtained from the Mikulski Archive for Space Telescopes (MAST) at STScI (\url{https://mast.stsci.edu}).

\begin{acknowledgements}
This work used TESS data from the Mikulski Archive for Space Telescopes (MAST) at STScI. Analysis was performed using Python~3.11 with \texttt{lightkurve}~2.4, \texttt{transitleastsquares}~1.0.31, \texttt{triceratops}~1.0.18, \texttt{celerite2}~0.3, \texttt{emcee}~3.1, and \texttt{batman}~2.4. The author thanks the development teams of these packages. The author also thanks Michael Hippke for endorsing the arXiv submission of this work.

Transit searches ran on Apple M-series and Intel Xeon CPUs ($\sim$800 CPU-hours total).

Regarding AI transparency: pipeline code was developed with assistance from Claude (Anthropic) for code generation and debugging, with cross-validation by Gemini (Google) for algorithm design review. All AI-generated code was reviewed, tested, and validated by the author. The manuscript text was written by the author; AI tools were used for language editing only. The full development history is preserved in the version control repository.

\end{acknowledgements}

%\FloatBarrier
%{\small
\bibliographystyle{aa}
\bibliography{references}

\begin{thebibliography}{56}
\expandafter\ifx\csname natexlab\endcsname\relax\def\natexlab#1{#1}\fi

\bibitem[{Barkaoui {et~al.}(2024)}]{Barkaoui2024}
Barkaoui, K. {et~al.} 2024, \aap, 687, A261

\bibitem[{Batalha {et~al.}(2013)}]{Batalha2013}
Batalha, N.~M. {et~al.} 2013, ApJS, 204, 24

\bibitem[{Bryson {et~al.}(2013)}]{Bryson2013}
Bryson, S.~T. {et~al.} 2013, PASP, 125, 889

\bibitem[{Chambers {et~al.}(1996)Chambers, Wetherill, \& Boss}]{Chambers1996}
Chambers, J.~E., Wetherill, G.~W., \& Boss, A.~P. 1996, Icarus, 119, 261

\bibitem[{Claret(2017)}]{Claret2017}
Claret, A. 2017, \aap, 600, A30

\bibitem[{Cooke {et~al.}(2021)Cooke, Pollacco, Lendl, {et~al.}}]{Cooke2021}
Cooke, B.~F., Pollacco, D., Lendl, M., {et~al.} 2021, MNRAS, 500, 5088

\bibitem[{Coughlin {et~al.}(2016)}]{Coughlin2016}
Coughlin, J.~L. {et~al.} 2016, ApJS, 224, 12

\bibitem[{{Crossfield} {et~al.}(2019){Crossfield}, {Waalkes}, {Newton},
  {et~al.}}]{Crossfield2019}
{Crossfield}, I. J.~M., {Waalkes}, W., {Newton}, E.~R., {et~al.} 2019, \apjl,
  883, L16

\bibitem[{Delrez {et~al.}(2022)}]{Delrez2022}
Delrez, L. {et~al.} 2022, A\&A, 667, A59

\bibitem[{D{\'e}vora-Pajares {et~al.}(2024)D{\'e}vora-Pajares, Pozuelos,
  {et~al.}}]{DevorasPajares2024}
D{\'e}vora-Pajares, M., Pozuelos, F.~J., {et~al.} 2024, MNRAS, 532, 4752

\bibitem[{Dholakia {et~al.}(2024)}]{Dholakia2024}
Dholakia, S. {et~al.} 2024, MNRAS, 531, 1276

\bibitem[{Dressing \& Charbonneau(2015)}]{Dressing2015}
Dressing, C.~D. \& Charbonneau, D. 2015, ApJ, 807, 45

\bibitem[{Fabricius {et~al.}(2021)}]{Fabricius2021}
Fabricius, C. {et~al.} 2021, \aap, 649, A5

\bibitem[{Fabrycky {et~al.}(2014)}]{Fabrycky2014}
Fabrycky, D.~C. {et~al.} 2014, ApJ, 790, 146

\bibitem[{Fausnaugh {et~al.}(2018)}]{Fausnaugh2018}
Fausnaugh, M.~M. {et~al.} 2018, {TESS} Data Release Notes: Sectors 1--69, Tech.
  Rep. NASA/TM-2018-220190, NASA Goddard Space Flight Center, updated per
  sector; documents 1-day Earth rotation systematic

\bibitem[{Foreman-Mackey {et~al.}(2013)}]{ForemanMackey2013}
Foreman-Mackey, D. {et~al.} 2013, PASP, 125, 306

\bibitem[{Foreman-Mackey {et~al.}(2017)}]{ForemanMackey2017}
Foreman-Mackey, D. {et~al.} 2017, AJ, 154, 220

\bibitem[{{Gaia Collaboration}(2023)}]{GaiaDR3}
{Gaia Collaboration}. 2023, \aap, 674, A1

\bibitem[{Giacalone \& Dressing(2020)}]{Giacalone2020}
Giacalone, S. \& Dressing, C.~D. 2020, AJ, 159, 169

\bibitem[{Giacalone {et~al.}(2021)}]{Giacalone2021}
Giacalone, S. {et~al.} 2021, AJ, 161, 24

\bibitem[{Gilbert {et~al.}(2020)}]{Gilbert2020}
Gilbert, E.~A. {et~al.} 2020, AJ, 160, 116

\bibitem[{Gillon {et~al.}(2017)Gillon, Triaud, Demory,
  {et~al.}}]{Gillon2017trappist}
Gillon, M., Triaud, A.~H.~M.~J., Demory, B.-O., {et~al.} 2017, Nature, 542, 456

\bibitem[{Gladman(1993)}]{Gladman1993}
Gladman, B. 1993, Icarus, 106, 247

\bibitem[{Guerrero {et~al.}(2021)}]{Guerrero2021}
Guerrero, N.~M. {et~al.} 2021, ApJS, 254, 39

\bibitem[{Henry {et~al.}(2006)}]{Henry2006}
Henry, T.~J. {et~al.} 2006, AJ, 132, 2360

\bibitem[{Hippke \& Heller(2019)}]{Hippke2019}
Hippke, M. \& Heller, R. 2019, \aap, 623, A39

\bibitem[{Hippke {et~al.}(2019)}]{Hippke2019wotan}
Hippke, M. {et~al.} 2019, AJ, 158, 143

\bibitem[{Jenkins {et~al.}(2002)Jenkins, Caldwell, \& Borucki}]{Jenkins2002}
Jenkins, J.~M., Caldwell, D.~A., \& Borucki, W.~J. 2002, ApJ, 564, 495

\bibitem[{Kemmer {et~al.}(2020)}]{Kemmer2020}
Kemmer, J. {et~al.} 2020, \aap, 642, A236

\bibitem[{Kempton {et~al.}(2018)}]{Kempton2018}
Kempton, E.~M.-R. {et~al.} 2018, PASP, 130, 114401

\bibitem[{Kostov {et~al.}(2019)}]{Kostov2019}
Kostov, V.~B. {et~al.} 2019, AJ, 158, 32

\bibitem[{Kreidberg(2015)}]{Kreidberg2015}
Kreidberg, L. 2015, PASP, 127, 1161

\bibitem[{Kunimoto {et~al.}(2025)Kunimoto, Shallue, Vanderburg, \&
  Gill}]{Kunimoto2025}
Kunimoto, M., Shallue, C.~J., Vanderburg, A., \& Gill, S. 2025, AJ, 169, 42

\bibitem[{{Lightkurve Collaboration}(2018)}]{Lightkurve2018}
{Lightkurve Collaboration}. 2018, Lightkurve: Kepler and TESS time series
  analysis in Python, Astrophysics Source Code Library, record ascl:1812.013

\bibitem[{Lindegren {et~al.}(2021)}]{Lindegren2021}
Lindegren, L. {et~al.} 2021, \aap, 649, A2

\bibitem[{Luque {et~al.}(2019)}]{Luque2019}
Luque, R. {et~al.} 2019, \aap, 628, A39

\bibitem[{Mann {et~al.}(2015)}]{Mann2015}
Mann, A.~W. {et~al.} 2015, ApJ, 804, 64

\bibitem[{Ment \& Charbonneau(2023)}]{Ment2023}
Ment, K. \& Charbonneau, D. 2023, AJ, 165, 265

\bibitem[{Newton {et~al.}(2017)Newton, Irwin, Charbonneau, Berlind, Calkins, \&
  Mink}]{Newton2017}
Newton, E.~R., Irwin, J., Charbonneau, D., {et~al.} 2017, ApJ, 834, 85

\bibitem[{Osborn {et~al.}(2016)Osborn, Armstrong, Brown, {et~al.}}]{Osborn2016}
Osborn, H.~P., Armstrong, D.~J., Brown, D.~J.~A., {et~al.} 2016, MNRAS, 457,
  2273

\bibitem[{{Peterson} {et~al.}(2023){Peterson}, {Benneke}, {Collins},
  {et~al.}}]{Peterson2023}
{Peterson}, M.~S., {Benneke}, B., {Collins}, K., {et~al.} 2023, \nat, 617, 701

\bibitem[{Plavchan {et~al.}(2020)}]{Plavchan2020}
Plavchan, P. {et~al.} 2020, Nature, 582, 497

\bibitem[{Pont {et~al.}(2006)Pont, Zucker, \& Queloz}]{Pont2006}
Pont, F., Zucker, S., \& Queloz, D. 2006, MNRAS, 373, 231

\bibitem[{Rao {et~al.}(2021)Rao, Mahabal, Rao, \& Raghavendra}]{Rao2021}
Rao, S., Mahabal, A., Rao, N., \& Raghavendra, C. 2021, MNRAS, 502, 2845

\bibitem[{Rauer {et~al.}(2024)Rauer, Catala, Aerts, {et~al.}}]{Rauer2024}
Rauer, H., Catala, C., Aerts, C., {et~al.} 2024, Experimental Astronomy, 58, 2

\bibitem[{Ricker {et~al.}(2015)}]{Ricker2015}
Ricker, G.~R. {et~al.} 2015, JATIS, 1, 014003

\bibitem[{Santerne {et~al.}(2013)}]{Santerne2013}
Santerne, A. {et~al.} 2013, \aap, 557, A139

\bibitem[{Sebastian {et~al.}(2021)Sebastian, Gillon, Ducrot,
  {et~al.}}]{Sebastian2021}
Sebastian, D., Gillon, M., Ducrot, E., {et~al.} 2021, A\&A, 645, A100

\bibitem[{Stassun {et~al.}(2019)}]{Stassun2019}
Stassun, K.~G. {et~al.} 2019, AJ, 158, 138

\bibitem[{Sullivan {et~al.}(2015)}]{Sullivan2015}
Sullivan, P.~W. {et~al.} 2015, ApJ, 809, 77

\bibitem[{Thompson {et~al.}(2018)}]{Thompson2018}
Thompson, S.~E. {et~al.} 2018, ApJS, 235, 38

\bibitem[{Timmer \& K{\"o}nig(1995)}]{Timmer1995}
Timmer, J. \& K{\"o}nig, M. 1995, A\&A, 300, 707

\bibitem[{Trifonov {et~al.}(2021)}]{Trifonov2021}
Trifonov, T. {et~al.} 2021, Science, 371, 1038

\bibitem[{Twicken {et~al.}(2018)}]{Twicken2018}
Twicken, J.~D. {et~al.} 2018, PASP, 130, 064502

\bibitem[{Yaptangco {et~al.}(2025)Yaptangco, Ballard, \&
  Dittmann}]{Yaptangco2025}
Yaptangco, G., Ballard, S., \& Dittmann, J. 2025, AJ, 169, 153

\bibitem[{Zechmeister \& K{\"u}rster(2009)}]{Zechmeister2009}
Zechmeister, M. \& K{\"u}rster, M. 2009, A\&A, 496, 577

\end{thebibliography}
%}

\begin{appendix}

\section{Complete candidate catalog}
\label{app:catalog}

\begin{table}[H]
\centering
\caption{Complete survey statistics.}
\label{tab:app_survey}
\scriptsize
\begin{tabular}{@{}lrl@{}}
\toprule
\textbf{Metric} & \textbf{Value} & \textbf{Notes} \\
\midrule
Stars analyzed & 121 & Complete sample \\
No detection & 98 & 81.0\% \\
With signals & 23 & 19.0\% \\
\midrule
Post-cleanup & 33 & Harmonic nest cleanup v1.1 \\
Surviving & 20 & 16 systems \\
Multi-planet & 8 cand. & 4 systems \\
TOI crossmatch & 0 & --- \\
\midrule
PLANET\_CAND & 9 / 8 sys & BEB excluded or LIKELY\_PLANET \\
NEEDS\_HR & 11 / 9 sys & HR imaging required \\
LIKELY\_EB & 9 / 8 sys & Eclipsing binary \\
FALSE\_POS & 3 / 3 sys & Contamination certain \\
GRAZING & 1 / 1 sys & Radius uncertain \\
\bottomrule
\end{tabular}
\tablefoot{Excludes 13 harmonic nests and LIKELY\_EB/FALSE\_POSITIVE signals from post-cleanup count.}
\end{table}

\subsection{PLANET\_CANDIDATE}

\begin{table*}[!t]
\centering
\caption{PLANET\_CANDIDATE: exhaustive parameters.}
\label{tab:app_pc}
\scriptsize
\begin{tabular}{lccccccccccccc}
\toprule
TIC & P\# & P (d) & $T_0$ (BTJD) & Depth (ppm) & SDE & $R_p$ ($R_\oplus$) & $T_{\mathrm{eq}}$ (K) & S (S$_\oplus$) & $T_{\rm mag}$ & d (pc) & RUWE & Tier & HZ \\
\midrule
117955697 & 1 & 1.833 & 2854.62 & 617 & 8.20 & $2.78^{+0.06}_{-0.08}$ & $605 \pm 55$ & 31.74 & 11.82 & 32.6 & 1.30 & 2 & \\
117955697 & 2 & 3.789 & 2855.08 & 877 & 7.96 & $3.12^{+0.06}_{-0.07}$ & $475 \pm 43$ & 12.05 & 11.82 & 32.6 & 1.30 & 2 & \\
77634567 & 1 & 4.123 & 1792.04 & 1164 & 8.04 & $0.99^{+0.14}_{-0.14}$ & $415 \pm 38$ & 7.03 & 12.36 & 30.7 & 1.33 & 2 & \\
417927999 & 1 & 1.574 & 1817.19 & 737 & 9.20 & $0.76^{+0.17}_{-0.16}$ & $609 \pm 55$ & 32.66 & 11.75 & 26.4 & 1.24 & 2 & \\
24293836 & 2 & 16.242 & 1328.05 & 1316 & 8.55 & $1.16^{+0.25}_{-0.24}$ & $284 \pm 26$ & 1.54 & 11.51 & 26.1 & 1.30$^\dagger$ & 3$^\dagger$ & \\
129766762 & 1 & 1.122 & 1386.15 & 626 & 8.03 & $0.84^{+0.08}_{-0.09}$ & $681 \pm 62$ & 51.01 & 12.22 & 33.6 & 1.31$^\dagger$ & 3 & \\
139471702 & 1 & 41.184 & 1398.45 & 2526 & 8.72 & $1.57^{+0.32}_{-0.31}$ & $205 \pm 19$ & 0.42 & 12.39 & 37.3 & 1.21 & 3 & \textbf{Yes} \\
80315364 & 1 & 22.122 & 2095.39 & 2173 & 7.84 & $0.79^{+0.17}_{-0.16}$ & $186 \pm 17$ & 0.28 & 12.37 & 14.6 & 1.37 & 3 & \textbf{Yes} \\
\midrule
72750190 & 1 & 50.340 & 1407.16 & 4978 & 9.60 & $2.01^{*}$ & $187 \pm 17$ & 0.29 & 12.14 & 31.1 & 1.15$^\dagger$ & M$^\S$ & Tent.$^\ddagger$ \\
\bottomrule
\end{tabular}
\tablefoot{Ordered by signal reliability tier (Sect.~\ref{sec:tier}); monotransit last. $R_p$ from MCMC unless noted. $T_{\mathrm{eq}}$ assumes Bond albedo $A=0.3$; uncertainties propagated from TIC~v8.2 stellar radii ($\sigma_{R_\star}/R_\star \approx 3$\%, combined with $T_{\mathrm{eff}}$ uncertainties, yielding $\sim$9\% on $T_{\mathrm{eq}}$). Tier=signal reliability tier (Sect.~\ref{sec:tier}): 1=high robustness, 2=moderate robustness, 3=noise-susceptible; M=monotransit (excluded from tier system). $^\dagger$LIKELY\_PLANET protected: Gaia contaminant(s) present or Gaia coverage incomplete, but TRICERATOPS FPP $< 50\%$; HR imaging recommended. $^*$TLS box-model estimate (MCMC did not converge). $^\ddagger$Tentative HZ detection: single transit epoch, MCMC failed; excluded from HZ candidate count. $^\S$Monotransit candidate: single transit epoch, excluded from tier system per \citet{Osborn2016, Cooke2021}.}
\end{table*}

\begin{table}[H]
\centering
\caption{Host star parameters for PLANET\_CANDIDATE systems.}
\label{tab:app_pc_stars}
\scriptsize
\begin{tabular}{@{}lccccccc@{}}
\toprule
TIC & SpT & $T_{\mathrm{eff}}$ & $R_\star$ & $M_\star$ & $T_{\rm mag}$ & d & Cand. \\
 & & (K) & ($R_\odot$) & ($M_\odot$) & & (pc) & \\
\midrule
80315364 & M5V & 2953 & 0.16 & 0.13 & 12.37 & 14.6 & 1 \\
77634567 & M3V & 3211 & 0.27 & 0.24 & 12.36 & 30.7 & 1 \\
72750190 & M3V & 3259 & 0.29 & 0.27 & 12.14 & 31.1 & 1 \\
129766762 & M3V & 3323 & 0.30 & 0.27 & 12.22 & 33.6 & 1 \\
139471702 & M3V & 3347 & 0.29 & 0.26 & 12.39 & 37.3 & 1 \\
24293836 & M3V & 3369 & 0.30 & 0.27 & 11.51 & 26.1 & 1$^a$ \\
417927999 & M3V & 3374 & 0.28 & 0.26 & 11.75 & 26.4 & 1 \\
117955697 & M3V & 3376 & 0.33 & 0.31 & 11.82 & 32.6 & 2 \\
\bottomrule
\end{tabular}
\tablefoot{Cand.\=number of candidates per system. $^a$+1 NEEDS\_HR\_IMAGING (P1).}
\end{table}

\subsection{NEEDS\_HR\_IMAGING}

\begin{table*}[!t]
\centering
\caption{NEEDS\_HR\_IMAGING: exhaustive parameters.}
\label{tab:app_hr}
\scriptsize
\begin{tabular}{lccccccccccc}
\toprule
TIC & P\# & P (d) & Depth (ppm) & SDE & $R_p$ ($R_\oplus$) & $T_{\mathrm{eq}}$ (K) & S (S$_\oplus$) & FPP (\%) & Contam. & Margin & HZ \\
\midrule
186664624 & 2 & 47.594 & 3995 & 8.07 & $2.31^{+0.47}_{-0.48}$ & $204 \pm 19$ & 0.41 & 60 & 2 & +0.4 & \textbf{Yes} \\
186664624 & 1 & 82.218 & 1497 & 8.38 & $1.43^{+0.29}_{-0.28}$ & $170 \pm 15$ & 0.20 & 77 & 3 & +1.5 & No \\
24293836 & 1 & 17.674 & 1145 & 8.25 & $1.72^{+0.13}_{-0.12}$ & $276 \pm 25$ & 1.37 & 53 & 2 & +1.2 & No \\
65516102 & 1 & 9.198 & 1236 & 9.45 & $1.00^{+0.19}_{-0.21}$ & $324 \pm 29$ & 2.62 & 66 & 1 & +1.4 & No \\
322053216 & 1 & 6.560 & 830 & 8.60 & $0.96^{+0.15}_{-0.16}$ & $376 \pm 34$ & 4.74 & 61 & 1 & +1.3 & No \\
124936003 & 1 & 1.156 & 952 & 10.71 & $0.69^{+0.15}_{-0.12}$ & $602 \pm 55$ & 31.21 & 67 & 2 & +3.8 & No \\
266523603 & 1 & 1.034 & 471 & 9.88 & $0.68^{+0.12}_{-0.11}$ & $732 \pm 67$ & 68.23 & 62 & 1 & +0.4 & No \\
123368775 & 1 & 0.727 & 358 & 9.79 & $0.71^{+0.10}_{-0.09}$ & $812 \pm 74$ & 103.20 & 58 & 1 & +1.6 & No \\
419939149 & 1 & 0.547 & 987 & 9.16 & $1.01^{+0.29}_{-0.19}$ & $901 \pm 82$ & 156.22 & 53 & 0$^*$ & --- & No \\
211401412 & 1 & 0.422 & 673 & 9.18 & $1.56^{+0.04}_{-0.03}$ & $931 \pm 85$ & 177.80 & 58 & 3$^*$ & +3.2 & No \\
211401412 & 2 & 0.629 & 888 & 7.81 & $1.69^{+0.05}_{-0.03}$ & $815 \pm 74$ & 104.38 & 57 & 3$^*$ & +2.9 & No \\
\bottomrule
\end{tabular}
\tablefoot{Contam.=number of Gaia sources with $G < G_{\rm max}$ within 30\arcsec. $R_p$ in $R_\oplus$ from MCMC. $T_{\mathrm{eq}}$ uncertainties as in Table~\ref{tab:app_pc}. Margin=brightest contaminant magnitude minus $G_{\rm max}$. $^*$Gaia incomplete at required $G_{\rm max}$.}
\end{table*}

\begin{table}[H]
\centering
\caption{HR imaging requirements and recommended facilities.}
\label{tab:app_hr_facilities}
\scriptsize
\begin{tabular}{@{}lcccl@{}}
\toprule
TIC & $G_{\rm max}$ & Marg. & Diff. & Facility \\
\midrule
186664624 & 18.5--19.5 & +0.4 & Tight & Gemini/`Alopeke \\
24293836 & 18.9 & +1.2 & Med. & SOAR/HRCam \\
65516102 & 19.7 & +1.4 & Easy & Any speckle \\
322053216 & 19.2 & +1.3 & Easy & Any speckle \\
124936003 & 19.8 & +3.8 & Easy & Any speckle \\
266523603 & 19.7 & +0.4 & Tight & Gemini/`Alopeke \\
123368775 & 20.3$^*$ & +1.6 & Med. & SOAR/HRCam \\
419939149 & 20.0$^*$ & N/A & Hard & Keck/NIRC2 \\
211401412 & 20.1--20.4$^*$ & +2.9 & Easy & Any speckle \\
\bottomrule
\end{tabular}
\tablefoot{Marg.=brightest contaminant $G - G_{\rm max}$. $^*$Gaia incomplete; deep AO required.}
\end{table}

\begin{figure*}[!t]
\centering
\includegraphics[width=0.32\textwidth,height=3.5cm]{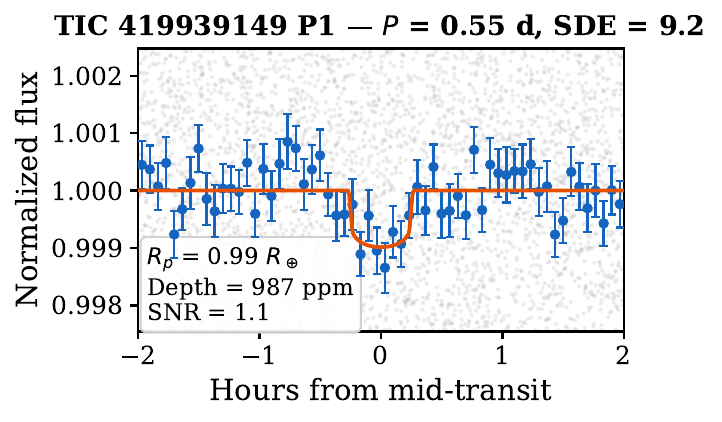}\hfill
\includegraphics[width=0.32\textwidth,height=3.5cm]{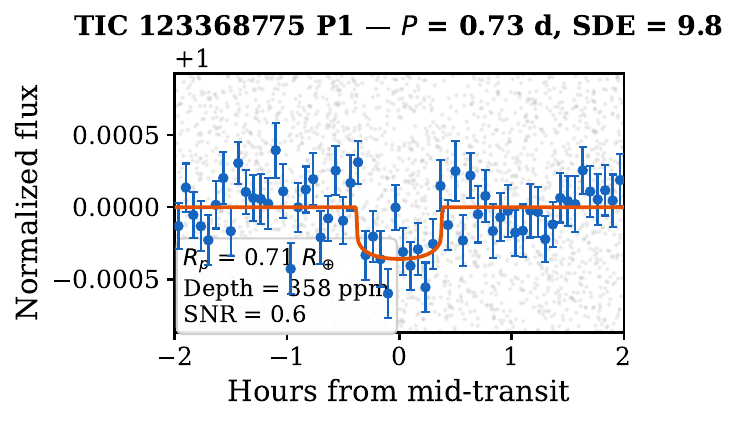}\hfill
\includegraphics[width=0.32\textwidth,height=3.5cm]{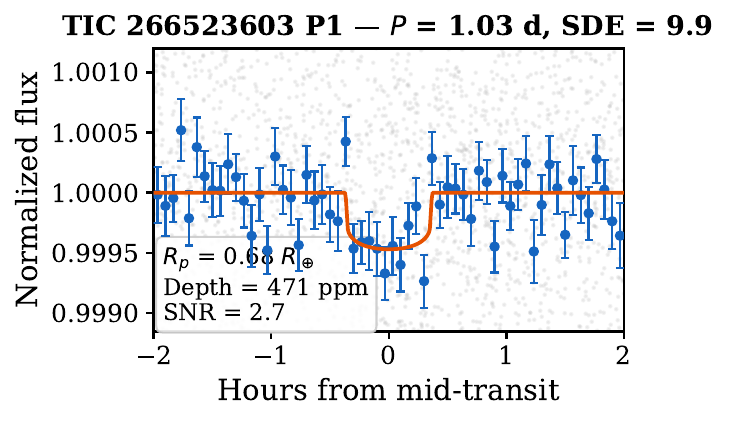}\\[2pt]
\includegraphics[width=0.32\textwidth,height=3.5cm]{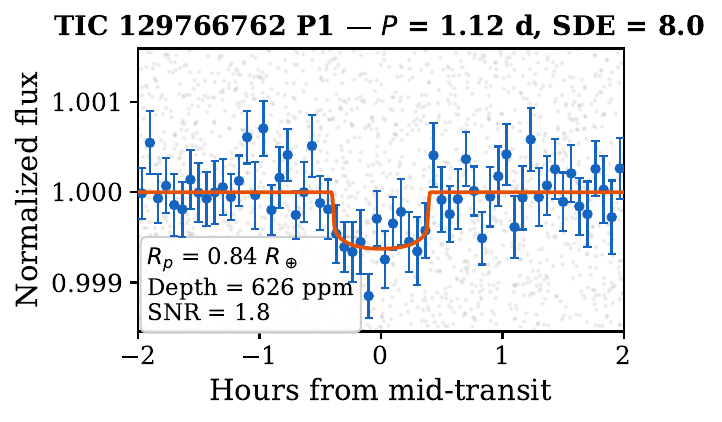}\hfill
\includegraphics[width=0.32\textwidth,height=3.5cm]{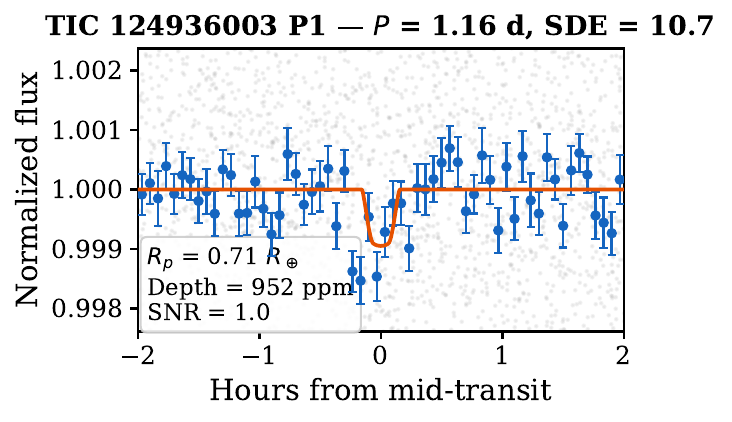}\hfill
\includegraphics[width=0.32\textwidth,height=3.5cm]{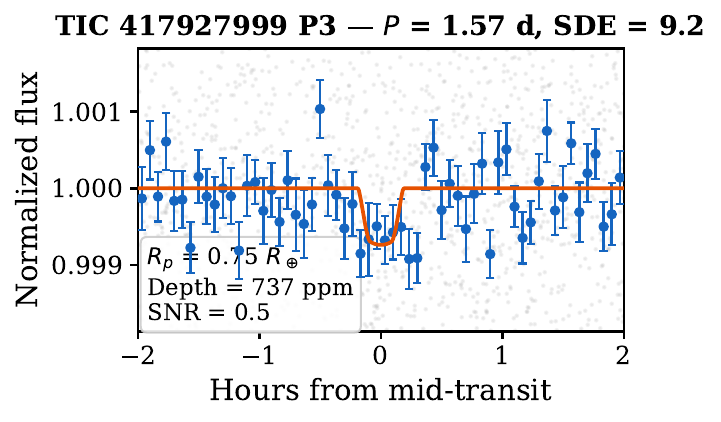}\\[2pt]
\includegraphics[width=0.32\textwidth,height=3.5cm]{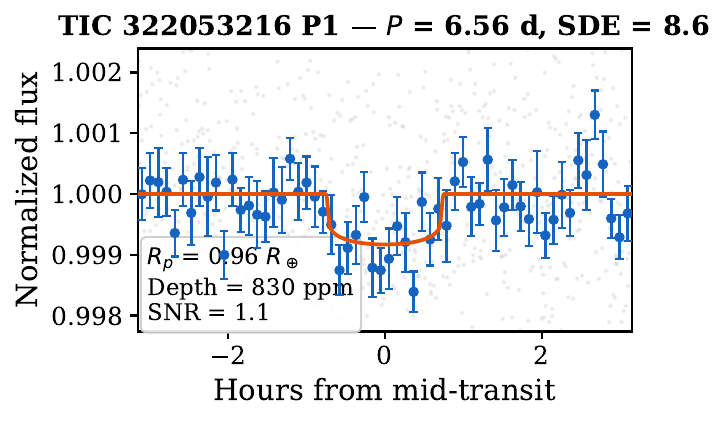}\hfill
\includegraphics[width=0.32\textwidth,height=3.5cm]{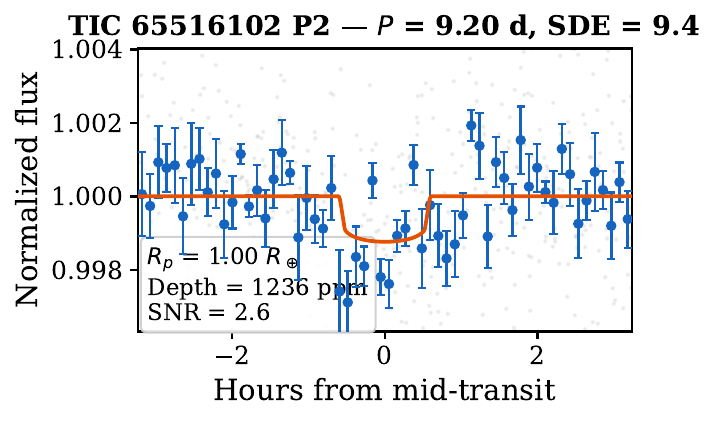}\hfill
\includegraphics[width=0.32\textwidth,height=3.5cm]{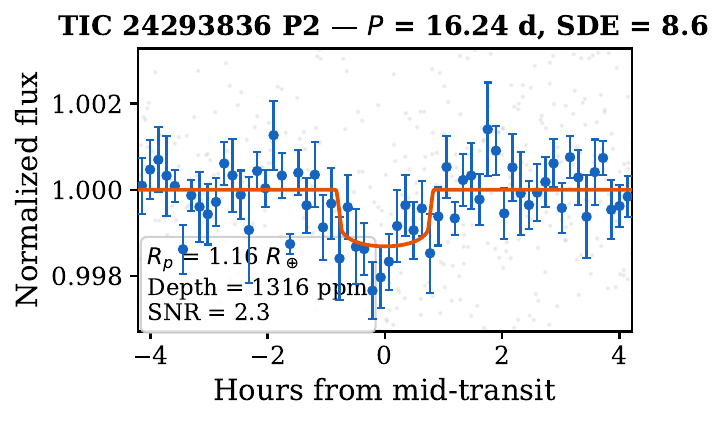}\\[2pt]
\includegraphics[width=0.32\textwidth,height=3.5cm]{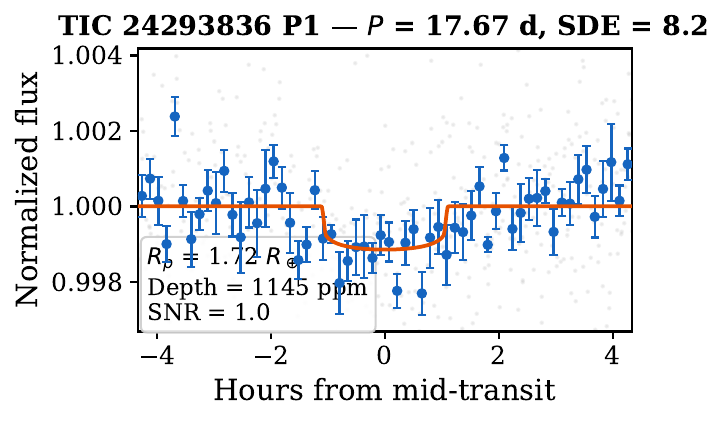}\hfill
\includegraphics[width=0.32\textwidth,height=3.5cm]{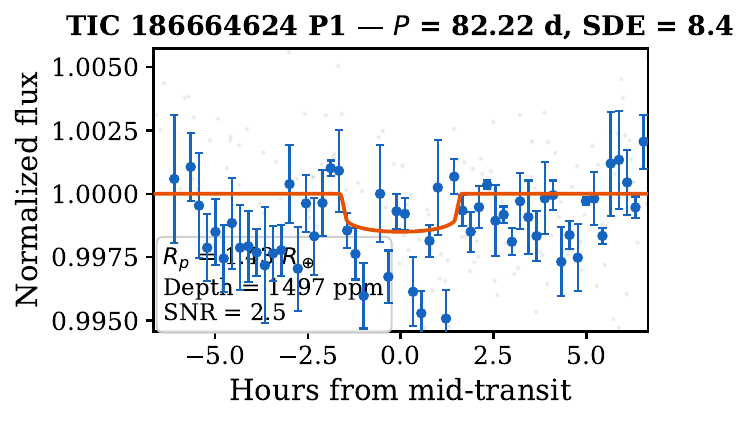}\hfill
\includegraphics[width=0.32\textwidth,height=3.5cm]{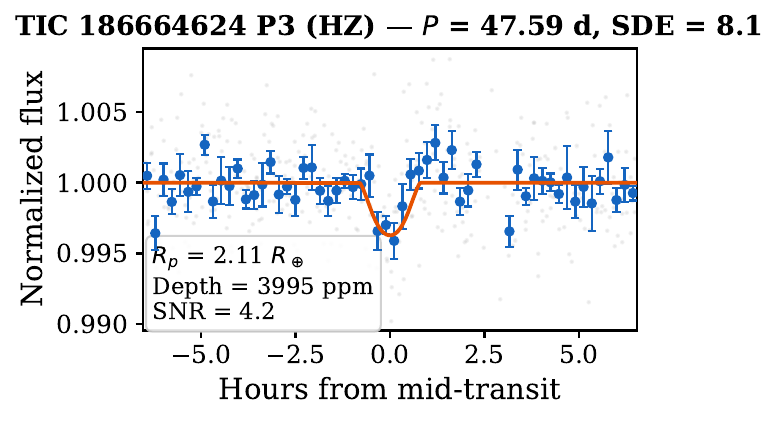}\\[2pt]
\includegraphics[width=0.32\textwidth,height=3.5cm]{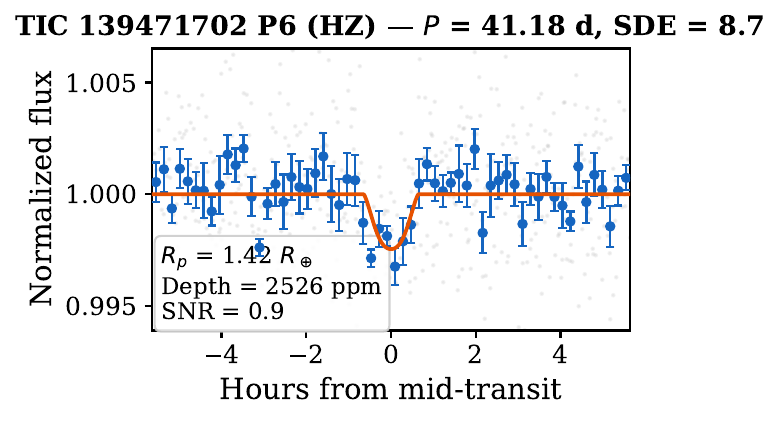}\hfill
\includegraphics[width=0.32\textwidth,height=3.5cm]{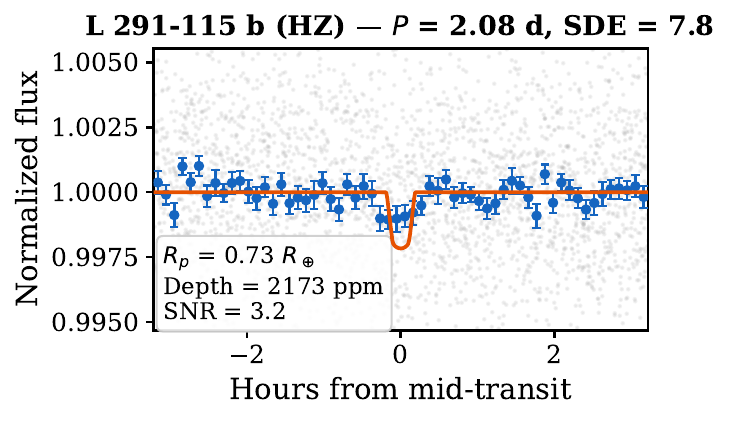}\hfill
\hspace{0.32\textwidth}
\caption{Phase-folded lightcurves for the 14 candidates not shown in Fig.~\ref{fig:phasefold_tier1} (\texttt{PLANET\_CANDIDATE} and \texttt{NEEDS\_HR\_IMAGING}), ordered by orbital period. Format as in Fig.~\ref{fig:phasefold_tier1}. Grey: individual measurements; blue: mean-binned data; orange: \texttt{batman} model with MCMC shape parameters. The three HZ candidates (TIC~186664624~P2, TIC~139471702, TIC-80315364; all tier~3) are included here. TIC~72750190 ($P=50.3$~d, tentative HZ, monotransit) is excluded: MCMC did not converge due to a single partial transit. Per-panel S/N values are per-transit; the phase-folded (combined) S/N equals $\sqrt{N_{\rm tr}/\beta_{\rm rn}}$ times the per-transit value (Sect.~\ref{sec:search}).}
\label{fig:phasefold_appendix}
\end{figure*}

\subsection{GRAZING\_CANDIDATE}

\begin{table}[H]
\centering
\caption{GRAZING\_CANDIDATE parameters.}
\label{tab:app_grazing}
\scriptsize
\begin{tabular}{@{}lccccc@{}}
\toprule
\textbf{TIC} & \textbf{P (d)} & \textbf{SDE} & \textbf{$R_p$$^*$} & \textbf{$b$} & \textbf{$T_{\mathrm{eq}}$} \\
\midrule
65516102 & 0.97 & 11.13 & $0.72$ & 1.07 & 728 \\
\bottomrule
\end{tabular}
\tablefoot{$^*$Radius ($R_\oplus$) highly uncertain: grazing geometry ($b > 1$).}
\end{table}

\subsection{LIKELY\_EB}

\begin{table}[H]
\centering
\caption{LIKELY\_EB candidates and rejection reasons.}
\label{tab:app_eb}
\scriptsize
\begin{tabular}{@{}lcccl@{}}
\toprule
TIC & P (d) & SDE & FPP & Reason \\
\midrule
206596030 & 97.27 & 10.2 & 28 & Dur.\ anomaly \\
206596030 & 50.31 & 11.4 & 45 & prob\_eb 25\% \\
85595350 & 80.30 & 8.7 & 60 & Dur.\ anomaly \\
53845811 & 67.56 & 12.6 & 56 & MCMC fail + EB \\
139471702 & 70.94 & 9.6 & 95 & NFPP=91\% \\
36727984 & 53.74 & 11.2 & 70 & Dur.\ anomaly \\
417927999 & 38.13 & 9.6 & 58 & prob\_eb 33\% \\
117955697 & 9.55 & 8.3 & 64 & 2nd eclipse \\
24293836 & 4.90 & 8.6 & 53 & 2nd eclipse \\
\bottomrule
\end{tabular}
\tablefoot{FPP in \%. Rejected signals are identified by orbital period; they do not carry planet numbers.}
\end{table}

\subsection{FALSE\_POSITIVE}

\begin{table}[H]
\centering
\caption{FALSE\_POSITIVE systems and rejection reasons.}
\label{tab:app_fp}
\scriptsize
\begin{tabular}{@{}lccl@{}}
\toprule
\textbf{TIC} & \textbf{P (d)} & \textbf{FPP} & \textbf{Reason} \\
\midrule
24292625 & 47.13 & 100\% & NFPP=100\% \\
123849672 & 75.92 & 100\% & NFPP=100\% \\
436558874 & 39.93 & 62\% & prob\_eb 39\% \\
\bottomrule
\end{tabular}
\end{table}

\subsection{Numerical vetting diagnostics for the 20 detected signals}
\label{app:vetting_numerics}

Table~\ref{tab:app_vetting_numerics} reports the numerical values measured per signal by the signal-validation cascade (Sect.~\ref{sec:signal_validation}) and by the TRICERATOPS and Gaia DR3 vetting steps (Sect.~\ref{sec:vetting}, Sect.~\ref{sec:physical_verification}). Each value is to be compared against the corresponding threshold of Table~\ref{tab:checks}; values within 20\% of a rejection boundary are marked with a dagger~($\dagger$) to highlight marginal passes. This granular view complements the pass/fail summary of Table~\ref{tab:classification} and is intended to assist follow-up planning by indicating which diagnostics sit close to their rejection thresholds for each candidate.

\begin{table*}[!t]
\centering
\caption{Numerical vetting diagnostics for the 20 detected signals.}
\label{tab:app_vetting_numerics}
\footnotesize
\setlength{\tabcolsep}{4pt}
\begin{tabular}{@{}lrrrrrrrrrr@{}}
\toprule
TIC-P\# & $P$ (d) & $N_{\rm tr}$ & SDE & S/N & $T_{\rm obs}/T_{\rm exp}$ & O/E $\sigma$ & Sec.\,$\delta$ & Cent.\,$\sigma$ & FPP (\%) & RUWE \\
\midrule
211401412-1 & 0.422 & 2649 & 9.2 & 0.6 & 1.02 & 1.71 & 0.203$^{\dagger}$ & 0.44 & 57.7 & -- \\
211401412-2 & 0.629 & 1776 & 7.8 & 0.8 & 1.01 & 1.86 & -0.182 & 1.18 & 56.7 & -- \\
417927999 & 1.574 & 1173 & 9.2 & 0.5 & 1.02 & 0.00 & 0.000 & 6.91$^{\dagger}$ & 35.6 & 1.24$^{\dagger}$ \\
117955697-1 & 1.833 & 413 & 8.2 & 1.7 & 1.02 & 0.00 & 0.000 & 4.63$^{\dagger}$ & -- & 1.30$^{\dagger}$ \\
117955697-2 & 3.789 & 200 & 8.0 & 1.5 & 1.02 & 0.00 & 0.000 & 3.20 & -- & 1.30$^{\dagger}$ \\
77634567 & 4.123 & 448 & 8.0 & 2.2 & 1.01 & 0.00 & 0.000 & 3.66 & 58.5 & 1.33$^{\dagger}$ \\
266523603 & 1.034 & 1812 & 9.9 & 2.7 & 1.02 & 0.31 & 0.029 & 4.99$^{\dagger}$ & 61.8 & -- \\
124936003 & 1.156 & 1922 & 10.7 & 1.0 & 1.01 & 0.00 & 0.000 & 1.72 & 67.3 & 1.32$^{\dagger}$ \\
322053216 & 6.560 & 393 & 8.6 & 1.1 & 1.02 & 0.28 & -0.117 & 3.25 & 60.9 & 1.39$^{\dagger}$ \\
24293836-2 & 16.242 & 154 & 8.5 & 2.3 & 1.01 & 0.80 & -0.206 & 14.72$^{\dagger}$ & -- & 1.30$^{\dagger}$ \\
129766762 & 1.122 & 2320 & 8.0 & 1.8 & 1.02 & 1.19 & 0.076 & 5.08$^{\dagger}$ & 40.7 & 1.31$^{\dagger}$ \\
139471702 & 41.184 & 63 & 8.7 & 0.9 & 1.00 & 0.00 & 0.000 & 3.34 & 60.0 & 1.21$^{\dagger}$ \\
80315364 & 22.122 & 84 & 7.8 & -0.8 & 1.00 & 0.00 & 0.000 & 4.10$^{\dagger}$ & 77.3 & 1.37$^{\dagger}$ \\
24293836-1 & 17.674 & 142 & 8.3 & 1.0 & 1.01 & 0.26 & -0.070 & 16.31$^{\dagger}$ & 53.5 & 1.30$^{\dagger}$ \\
419939149 & 0.547 & 2526 & 9.2 & 1.1 & 1.01 & 0.65 & -0.073 & 3.43 & 52.7 & 1.18$^{\dagger}$ \\
123368775 & 0.727 & 3577 & 9.8 & 0.6 & 1.03 & 1.84 & 0.068 & 0.74 & 57.6 & 1.14$^{\dagger}$ \\
65516102 & 9.198 & 280 & 9.4 & 2.6 & 1.01 & 0.56 & -0.438 & 4.96$^{\dagger}$ & 66.2 & 1.28$^{\dagger}$ \\
186664624-2 & 47.594 & 39 & 8.1 & 4.2 & 0.98 & 0.00 & 0.000 & 34.73$^{\dagger}$ & 60.0 & 1.25$^{\dagger}$ \\
186664624-1 & 82.218 & 23 & 8.4 & 2.5 & 0.25$^{\dagger}$ & 0.00 & 0.000 & 27.55$^{\dagger}$ & 76.9 & 1.25$^{\dagger}$ \\
72750190 & 50.340 & 52 & 9.6 & 2.5 & 0.97 & 0.00 & -0.079 & -- & -- & -- \\
\bottomrule
\end{tabular}
\tablefoot{Numerical values measured per signal against the thresholds of Table~\ref{tab:checks}. $N_{\rm tr}$: number of observed transits; $T_{\rm obs}/T_{\rm exp}$: duration plausibility ratio (threshold $\in [0.3, 3.0]$); O/E $\sigma$: odd-even depth difference (threshold $< 3\sigma$); Sec.\,$\delta$: TLS-reported secondary eclipse depth ratio at phase 0.5 (informational). The cascade rejection (check 7) uses a dedicated per-epoch secondary search with an S/N-gated rule (Sect.~\ref{sec:signal_validation}); the tabulated TLS ratio is not identical to the decision variable and can exceed 0.10 without triggering rejection when the secondary is not consistently detected across epochs. Cent.\,$\sigma$: centroid offset significance, reported for the diagnostic method selected by the pipeline (threshold $< 5\sigma$); FPP: TRICERATOPS false-positive probability, with "--" indicating TRICERATOPS server failure (TRILEGAL); RUWE: Gaia DR3 single-star indicator (threshold $< 1.4$). Values within 20\% of a rejection boundary are marked $\dagger$. Cascade checks returning a binary pass/fail outcome (SWEET variability, OOT scatter gate, detrend response) are satisfied for all 20 signals by construction of the cascade. TIC~80315364 diagnostics are derived from a dedicated vetting run on the P=22.12d signal preserved in a pre-harmonic-resolution pipeline state (see Sect.~\ref{sec:limitations}).}
\end{table*}

\subsection{Whole-sample light-curve inversion statistics}
\label{app:inversion_whole}

\begin{figure*}[!t]
\centering
\includegraphics[width=\textwidth]{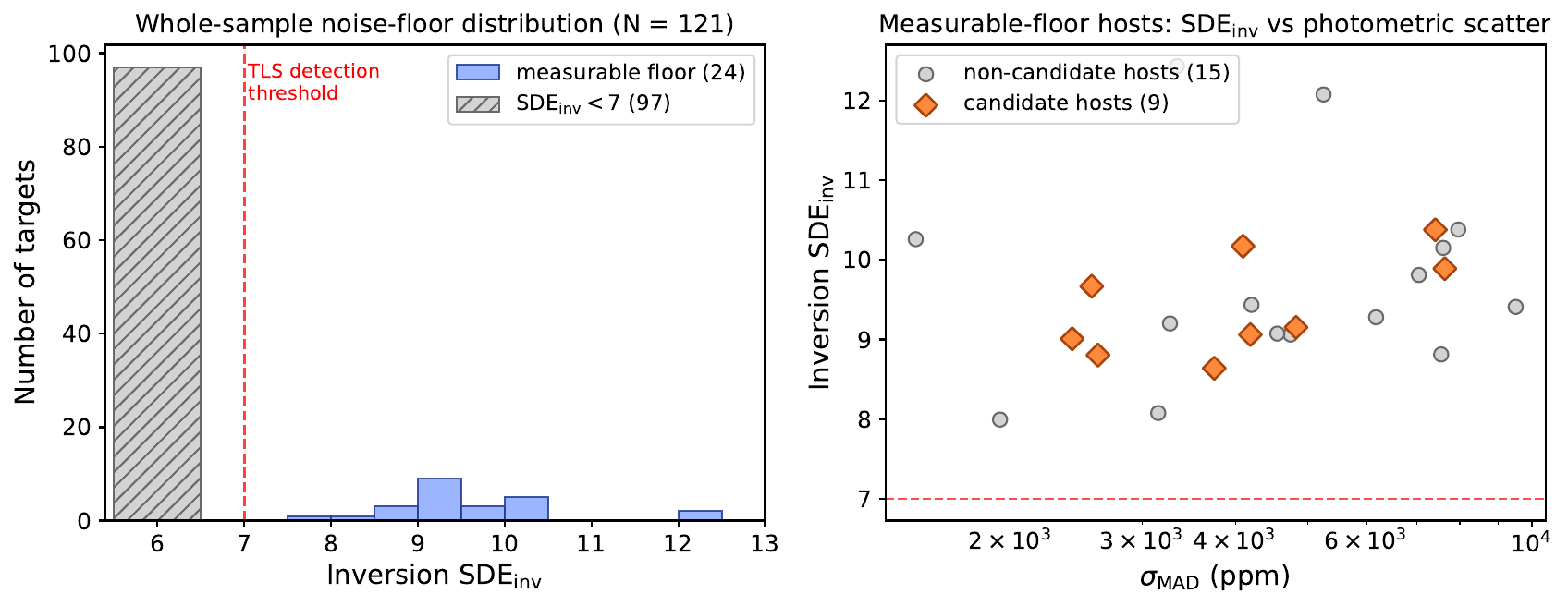}
\caption{Whole-sample light-curve inversion statistics (counts and discussion in Sect.~\ref{app:inversion_whole}). Left: Distribution of $\mathrm{SDE}_{\rm inv}$ across the 121 targets. Hatched grey bar at $\mathrm{SDE}_{\rm inv} < 7$: hosts with unresolved noise floor (BELOW). Blue histogram: Hosts with a measurable floor; red dashed line: TLS detection threshold. Right: $\mathrm{SDE}_{\rm inv}$ versus $\sigma_{\rm MAD}$ for the measurable-floor subset. Orange diamonds: Candidate hosts (COMP and DOM). Grey circles: Non-candidate hosts.}
\label{fig:inversion_whole}
\end{figure*}

The light-curve inversion procedure (Sect.~\ref{sec:far}) was applied to all 121 targets of the survey, extending the per-candidate diagnostic of Sect.~\ref{sec:far} into a population-level characterization of the noise floor. Of the 121 targets, 97 (80.2\%) yield $\mathrm{SDE}_{\rm inv} < 7$ and have noise floors that lie below the TLS detection threshold; the remaining 24 (19.8\%) reach $\mathrm{SDE}_{\rm inv}$ between 7.0 and 12.4 and provide a measurable per-star floor. Figure~\ref{fig:inversion_whole} (left) shows the resulting distribution. The 16 candidate hosts split as 7 BELOW, 5 COMP, and 4 DOM; their classifications relative to $\mathrm{SDE}_{\rm real}$ (Sect.~\ref{sec:far}) are recovered identically by the whole-sample run, confirming the per-candidate analysis.

The right panel of Fig.~\ref{fig:inversion_whole} plots $\mathrm{SDE}_{\rm inv}$ against $\sigma_{\rm MAD}$ for the 24 measurable-floor targets. The 9 candidate hosts among them (orange diamonds) are not preferentially located at the high-$\sigma_{\rm MAD}$ end: the noise-frontier classification is driven by the joint statistics of variability amplitude and detected-signal SDE rather than by photometric scatter alone, consistent with the tier-3 distribution reported in Sect.~\ref{sec:tier}. The 15 measurable-floor non-candidate hosts illustrate a regime where pure stellar variability drives the TLS to $\mathrm{SDE}_{\rm inv} \geq 7$ on inverted data: on such hosts, the SDE~$\geq 7$ threshold alone cannot distinguish genuine transits from variability peaks, and the combined FAR/inversion/Fourier framework is required for reliable candidate identification.

\FloatBarrier
\onecolumn

\section{Target sample}
\label{app:sample}

\noindent
\begin{minipage}[t]{0.48\textwidth}
\begin{table}[H]
\centering
\caption{M3--M6 sample summary statistics.}
\label{tab:sample_summary}
\begin{tabular}{lcc}
\toprule
\textbf{Parameter} & \textbf{Range} & \textbf{Median} \\
\midrule
$T_{\mathrm{eff}}$ (K) & 2848--3397 & 3280 \\
$T_{\rm mag}$ & 8.34--12.50 & 11.97 \\
$R_\star$ ($R_\odot$) & 0.147--0.349 & 0.280 \\
$M_\star$ ($M_\odot$) & 0.120--0.333 & 0.256 \\
RUWE & 1.00--1.39 & 1.05 \\
Distance (pc) & 5.2--48.2 & 26.5 \\
N sectors & 3--10 & 4 \\
Baseline (d) & 735--2602 & 1847 \\
\bottomrule
\end{tabular}
\end{table}
\end{minipage}\hfill
\begin{minipage}[t]{0.48\textwidth}
\begin{table}[H]
\centering
\caption{Sample distribution by spectral type and detection outcome.}
\label{tab:sample_dist}
\begin{tabular}{lcccc}
\toprule
\textbf{Type} & \textbf{$T_{\mathrm{eff}}$ (K)} & \textbf{N} & \textbf{Signals} & \textbf{Det.\ rate} \\
\midrule
M6 & $<$2900 & 2 & 0 & 0\% \\
M4--M5 & 2900--3200 & 22 & 6 & 27.3\% \\
M3 & 3200--3400 & 97 & 17 & 17.5\% \\
\midrule
\textbf{Total} & 2848--3397 & \textbf{121} & \textbf{23} & \textbf{19.0\%} \\
\bottomrule
\end{tabular}
\end{table}
\end{minipage}

\section{Validation results}
\label{app:validation}

\begin{table}[H]
\centering
\caption{Complete validation sample.}
\label{tab:app_val_overview}
\begin{tabular}{llccccccc}
\toprule
\textbf{System} & \textbf{TIC} & \textbf{SpT} & \textbf{d (pc)} & \textbf{$T_{\rm mag}$} & \textbf{Planets} & \textbf{HZ} & \textbf{Gaia} & \textbf{Status} \\
\midrule
L~98-59 & 307210830 & M3V & 10.6 & 9.39 & 3 & --- & 0 & PLANET\_CAND \\
TOI-700 & 150428135 & M2V & 31.1 & 10.95 & 4 & d & 2--5 & NEEDS\_HR \\
Gliese~12 & 52005579 & M4V & 12.2 & 10.18 & 1 & b & 0 & PLANET\_CAND \\
GJ~357 & 413248763 & M2.5V & 9.4 & 8.74 & 1 & --- & 2 & NEEDS\_HR \\
GJ~3473 & 452866790 & M4V & 27.4 & 11.20 & 1 & --- & 0 & PLANET\_CAND \\
GJ~486 & 390651552 & M3.5V & 8.1 & 8.07 & 1 & --- & 0 & PLANET\_CAND \\
TOI-406 & 153065527 & M4V & 22.0 & 11.28 & 1 & --- & 1 & NEEDS\_HR \\
TOI-782 & 429358906 & M3V & 57.0 & 12.29 & 1 & --- & 1 & NEEDS\_HR \\
TOI-6086 & 18318288 & M3.5V & 32.0 & 12.41 & 1 & --- & 1 & NEEDS\_HR \\
LP~791-18 & 181804752 & M5V & 26.5 & 13.56 & 2$^\dagger$ & --- & 1--2 & NEEDS\_HR \\
\midrule
\textbf{Total} & & & & & \textbf{16} & \textbf{2} & & \textbf{6 PC / 10 HR} \\
\bottomrule
\end{tabular}
\tablefoot{10 systems meeting criteria, 16 TESS-detectable planets, 100\% recovery. LHS~1140 tested separately as limit case. Gaia=contaminants within 30$''$ brighter than $R_{p,\rm max}$. $^\dagger$LP~791-18~d excluded (Spitzer discovery, \citealt{Peterson2023}).}
\end{table}

\begin{table}[H]
\centering
\caption{Validation planet detection parameters and NASA Archive comparison.}
\label{tab:app_val_detection}
\small
\begin{tabular}{llcccccccc}
\toprule
\textbf{System} & \textbf{Pl.} & \textbf{$P_{\rm det}$ (d)} & \textbf{$P_{\rm arch}$ (d)} & \textbf{$\Delta P/P$}$^a$ & \textbf{SDE} & \textbf{Depth} & \textbf{$R_p$} & \textbf{FPP} & \textbf{NFPP} \\
\midrule
L~98-59 & b & 2.2531 & 2.2531 & $<$0.01\% & 56.3 & 596 & 0.80 & 33\% & 26\% \\
L~98-59 & c & 3.6908 & 3.6907 & $<$0.01\% & 54.9 & 1656 & 1.38 & 0.6\% & 0.6\% \\
L~98-59 & d & 7.3810 & 7.4507 & 0.94\%$^b$ & 65.1 & 1279 & 1.49 & 2.9\% & 2.5\% \\
TOI-700 & b & 9.9769 & 9.9770 & $<$0.01\% & 37.6 & 453 & 1.01 & 29\% & 20\% \\
TOI-700 & c & 16.051 & 16.051 & $<$0.01\% & 49.4 & 2079 & 2.63 & 33\% & 16\% \\
TOI-700 & d & 37.426 & 37.426 & $<$0.01\% & 28.2 & 575 & 1.14 & 25\% & 15\% \\
TOI-700 & e & 27.811 & 27.809 & $<$0.01\% & 18.5 & 388 & 0.95 & 43\% & 22\% \\
Gliese~12 & b & 12.761 & 12.761 & $<$0.01\% & 23.5 & 1047 & 1.00 & 35\% & 17\% \\
GJ~357 & b & 3.9306 & 3.9306 & $<$0.01\% & 48.5 & 1023 & 1.26 & 20\% & 13\% \\
GJ~3473 & b & 1.1980 & 1.1980 & $<$0.01\% & 37.9 & 1239 & 1.51 & 21\% & 17\% \\
GJ~486 & b & 1.4671 & 1.4671 & $<$0.01\% & 27.9 & 1705 & 1.31 & 18\% & 14\% \\
TOI-406 & b & 13.176 & 13.176 & $<$0.01\% & 30.4 & 2255 & 1.88 & 16\% & 7\% \\
TOI-782 & b & 8.0240 & 8.0240 & $<$0.01\% & 36.4 & 3549 & 2.48 & 28\% & 20\% \\
TOI-6086 & b & 1.3889 & 1.3889 & $<$0.01\% & 32.4 & 1705 & 0.95 & 61\% & 57\% \\
LP~791-18 & c & 4.9899 & 4.9899 & $<$0.01\% & 46.1 & 17086 & 2.13 & 39\% & 28\% \\
LP~791-18 & b & 0.9480 & 0.9480 & $<$0.01\% & 42.9 & 4023 & 1.12 & 61\% & 53\% \\
\bottomrule
\end{tabular}
\tablefoot{Depth in ppm. $R_p$ in $R_\oplus$. FPP/NFPP from TRICERATOPS. $^a\Delta P/P=|P_{\rm det} - P_{\rm arch}| / P_{\rm arch}$. $^b$L~98-59~d is detected at $P=7.381$~d rather than the archive value of 7.451~d (0.94\% offset); this exceeds the 0.2\% injection-recovery tolerance (Sect.~\ref{sec:injection_method}) but is unambiguously the correct planet (SDE~$= 65.1$, highest in the system). The period offset likely results from iterative masking of planets b and c prior to the search for planet d.}
\end{table}

\begin{table}[H]
\centering
\caption{Gaia DR3 verification for validation planets.}
\label{tab:app_val_gaia}
\scriptsize
\begin{tabular}{@{}llcccl@{}}
\toprule
System & Pl. & $G_{\rm max}$ & Ct. & RUWE & Status \\
\midrule
L~98-59 & b & 17.5 & 0 & 1.27 & PC \\
L~98-59 & c & 16.4 & 0 & 1.27 & PC \\
L~98-59 & d & 16.6 & 0 & 1.27 & PC \\
TOI-700 & b & 19.3 & 5 & 1.03 & HR \\
TOI-700 & c & 17.6 & 2 & 1.03 & HR \\
TOI-700 & d & 19.0 & 5 & 1.03 & HR \\
TOI-700 & e & 19.4 & 5 & 1.03 & HR \\
Gliese~12 & b & 17.6 & 0 & 1.20 & PC \\
GJ~357 & b & 16.2 & 2 & 1.12 & HR \\
GJ~3473 & b & 18.5 & 0 & 1.30 & PC \\
GJ~486 & b & 16.8 & 0 & 1.09 & PC \\
TOI-406 & b & 17.9 & 1 & 1.26 & HR \\
TOI-782 & b & 18.4 & 1 & 1.25 & HR \\
TOI-6086 & b & 19.3 & 1 & 1.20 & HR \\
LP~791-18 & c & 18.0 & 1 & 1.12 & HR \\
LP~791-18 & b & 19.5 & 2 & 1.12 & HR \\
\midrule
LHS~1140$^\dagger$ & b & 16.8 & 1 & \textbf{1.53} & HR \\
LHS~1140$^\dagger$ & c & 17.5 & 1 & \textbf{1.53} & HR \\
\bottomrule
\end{tabular}
\tablefoot{Ct.=Gaia contaminants ($G < G_{\rm max}$, $<30''$). PC=PLANET\_CANDIDATE, HR=NEEDS\_HR\_IMAGING. $^\dagger$Limit test (2 sectors, RUWE $> 1.4$).}
\end{table}

\end{appendix}

\end{document}